%% file: Main.tex
\documentclass[
    reprint,
    amsmath,
    amssymb,
    aps,
    prd,
    floatfix,
    superscriptaddress 
]{revtex4-2}

\usepackage{graphicx}
\usepackage{dcolumn}
\usepackage{bm}
\usepackage{hyperref}
\usepackage[mathlines]{lineno}
\usepackage{gensymb}
\usepackage{xspace}
\usepackage{multirow}
\usepackage{xcolor}

\newcommand{\GEANTfour} {{\textsc{Geant4}}\xspace}
\newcommand{\GEANTfourRW} {{\textsc{Geant4reweight}}\xspace}
\newcommand{\GENIE} {{\textsc{Genie}}\xspace}
\newcommand{\CORSIKA} {{\textsc{Corsika}}\xspace}
\newcommand{\GUNDAM} {{\textsc{Gundam}}\xspace}
\newcommand{\NUWRO} {{\textsc{NuWro}}\xspace}
\newcommand{\GIBBU} {GiBBU\xspace}
\newcommand{\NEUT} {{\textsc{Neut}}\xspace}

\begin{document}

\preprint{APS/123-QED}

\title{Measurement of muon (anti-)neutrino charged-current quasielastic-like cross section using off-axis NuMI beam at ICARUS}


\input{Sections/author_institutions}
\collaboration{ICARUS Collaboration}
\noaffiliation

\date{\today}

\begin{abstract}
This paper presents the first neutrino cross-section measurement from the ICARUS detector at Fermilab, using NuMI (Neutrinos at the Main Injector) beam data collected from two beam operation periods corresponding to $2.5\times10^{20}$ protons-on-target in neutrino beam mode.
The signal is defined by events with no pions produced in the final state, a topology dominated by charged-current quasielastic-like (CCQE-like) signatures. 
The measurement is reported as flux-averaged differential cross sections as functions of kinematic variables that provide sensitivity to the complex nuclear effects which often dominate the systematic uncertainty budgets of neutrino oscillation measurements.
Specifically, this work reports cross sections in two angular variables---the angle of the outgoing lepton and the opening angle between the lepton and leading proton---and two variables characterizing the kinematic imbalance between the muon and proton in the plane transverse to the incoming neutrino.
These results are compared against predictions from a variety of neutrino event generators, with $p$-values calculated between the extracted cross sections and each prediction.
Overall, the predictions agree with the data; however, the current budget of uncertainties does not yet provide sufficient discriminating power to favor a specific model.
\end{abstract}

\maketitle

\input{Sections/Introduction}
\input{Sections/ICARUSExperiment}
\input{Sections/EventSimulation}
\input{Sections/EventRecoAndSelection}
\input{Sections/Systematics}
\input{Sections/CrossSectionExtraction}
\input{Sections/Results}
\input{Sections/Conclusion}

\begin{acknowledgments}
This document was prepared by the ICARUS Collaboration using the resources of the Fermi National Accelerator Laboratory (Fermilab), a U.S. Department of Energy, Office of Science, HEP User Facility. Fermilab is managed by FermiForward Discovery Group, LLC, acting under Contract No. 89243024CSC000002.

This work was also supported by Istituto Nazionale di Fisica Nucleare (INFN, Italy), EU Horizon 2020 Research and Innovation Program under the Marie Sklodowska-Curie Grant Agreement Nos. 822185, 858199, and 101003460 and Horizon Europe Program research and innovation programme under the Marie Sklodowska-Curie Grant Agreement No. 101081478, the research contract per Law 240/2010, Art. 24 (3)(a), and D.G.R. 693/2023 (REF. PA:2023-20090/RER—CUP:J19J23000730002) by FSE+ 2021–2027. Furthermore the support of CERN in the detector overhauling within the Neutrino Platform framework, in the detector installation and commissioning, is acknowledged.
This work was also supported by the Anusandhan National Research Foundation (ANRF, India) under the Ramanujan Fellowship (Grant No. RJF/2025/000203).

The ICARUS Collaboration would like to thank the MINOS Collaboration for having provided the Side CRT panels as well as Double Chooz (University of Chicago) for the Bottom CRT panels. 
We also acknowledge the contribution of many SBND colleagues, in particular for the development of a number of simulation, reconstruction and analysis tools which are shared within the SBN program.
\end{acknowledgments}
\appendix

\input{Sections/Appendix_GUNDAM_Exraction}
\input{Sections/Appendix_pvalue}

\input{Sections/Appendix_LCurve}
\input{Sections/Appendix_IndvFitComparison}
\input{Sections/Appendix_XsecFineBins}


\bibliographystyle{apsrev4-2}
\bibliography{main}

\end{document}

%% file: Sections/author_institutions.tex
\author{F. Abd Alrahman}\altaffiliation{}\affiliation{University of Houston, Houston, TX 77204, USA}
\author{P. Abratenko}\altaffiliation{}\affiliation{Tufts University, Medford, MA 02155, USA}
\author{N. Abrego-Martinez}\altaffiliation{}\affiliation{Centro de Investigacion y de Estudios Avanzados del IPN (Cinvestav), Mexico City}
\author{A. Aduszkiewicz}\altaffiliation{}\affiliation{University of Houston, Houston, TX 77204, USA}
\author{F. Akbar}\altaffiliation{}\affiliation{University of Rochester, Rochester, NY 14627, USA}
\author{L. Aliaga Soplin}\altaffiliation{}\affiliation{University of Texas at Arlington, Arlington, TX 76019, USA}
\author{M. Artero Pons}\altaffiliation{}\affiliation{INFN Sezione di Padova and University, Padova, Italy}
\author{J. Asaadi}\altaffiliation{}\affiliation{University of Texas at Arlington, Arlington, TX 76019, USA}
\author{W. F. Badgett}\altaffiliation{}\affiliation{Fermi National Accelerator Laboratory, Batavia, IL 60510, USA}
\author{B. Baibussinov}\altaffiliation{}\affiliation{INFN Sezione di Padova and University, Padova, Italy}
% ...

%% file: Sections/Introduction.tex
\section{Introduction}
\label{sec:Intro}

The primary goal of current and future long-baseline neutrino oscillation experiments, such as DUNE~\cite{DUNE:2020ypp}, is to perform high-precision measurements of neutrino oscillation parameters including the mass ordering and possible CP violation in the leptonic sector~\cite{PDG}.
Achieving these goals requires an unprecedented level of control over systematics uncertainties,
which are currently dominated by the modeling of neutrino-nucleus interactions~\cite{T2K:2023smv, NOvA:2018gge}.
Some of the largest uncertainties come from the modeling of the nuclear physics processes (``nuclear effects'') which distort neutrino nucleus interactions with respect to their neutrino-nucleon interaction counterparts.
Nuclear effects modify both the final-state particle kinematics and content, altering the rate of detection of any given interaction channel as a function of all kinematic variables~\cite{NuSTEC:2017hzk}. 
For example, the modeling of the struck nucleons' initial state affects the final-state particle kinematics, while final-state interactions (FSI) affect both kinematics and the particle content of the final state, as particles can be re-scattered or absorbed in the nucleus. 
As a result, a sample of events with only a lepton and nucleons in the final state may have originated from an interaction channel that produced other hadrons at the vertex.
While a complete theoretical description would treat the vertex interaction and the subsequent propagation of hadrons as a single, quantum-mechanically coherent process, most event generators rely on a simplified factorization scheme~\cite{Nikolakopoulos:2022qkq}.
Recently, however, new frameworks have attempted to describe both initial and final states within a consistent theoretical description ~\cite{Gonzalez_Jimenez_2020,PhysRevC.105.025502,edrmf}.
Measurements of neutrino-nucleus cross sections as a function of variables sensitive to nuclear effects are therefore essential to validate various model predictions.

In this paper, we report a flux-averaged differential muon-neutrino cross-section measurement on an argon target using the Neutrinos at the Main Injector (NuMI) beam at the ICARUS detector.
The signal is defined by charged-current (CC) interactions with no pions in the final state,
targeting a quasielastic-like signature (CCQE-like).
We report the measurement as a function of four kinematic variables.
Two angular variables are chosen: the muon angle with respect to the neutrino direction ($\cos{\theta_\mu}$)  and the opening angle between the muon and the leading proton ($\cos{\theta_{\mu,p}}$).
These two angular variables can be reconstructed precisely even for the candidate tracks that exit the detector.
Two transverse kinetic imbalance (TKI) variables~\cite{PhysRevD.101.092001,PhysRevC.94.015503}, $\delta p_{T}$ and $\delta \alpha_{T}$, are calculated using momenta of the muon and the leading proton. The former is the summation of the two momenta projected onto the plane perpendicular to the neutrino direction.
The latter is the angle between the vector of $\delta p_{T}$ and the momentum transfer vector in the transverse plane.
The TKI variables can only be reasonably reconstructed when the momenta are well measured. A subset of events where muon candidates are contained inside the detector (Sec.~\ref{sec:RecoSel:EvSel}) are used when TKI variables are examined, such that the momentum could be measured using the methods discussed in Sec.~\ref{sec:RecoSel:Reco}.
These variables are chosen for their unique sensitivity to nuclear effects such as initial-state nucleon momentum distributions and FSI.
This measurement utilizes the off-axis NuMI beam that provides a broad range of neutrino energy spectrum up to few GeV, which are complementary to the similar measurements performed on argon targets at lower~\cite{PhysRevD.102.112013,ubCC0pi2025} and higher~\cite{PhysRevLett.108.161802} energies.

The paper is organized as follows: Section~\ref{sec:ICARUS} describes the ICARUS detector and the NuMI beam; Section~\ref{sec:eventsim} details the simulation and modeling frameworks; Section~\ref{sec:RecoSel} discusses the event selection and signal definition; Section~\ref{sec:Systematics} outlines the systematic uncertainties; Section~\ref{sec:extraction} presents the cross-section extraction procedure; the results and their comparison to model predictions are shown in Section~\ref{sec:ResultAndComparison}; and Section~\ref{sec:concl} summarizes the analysis.

%% file: Sections/ICARUSExperiment.tex
\section{The ICARUS experiment}
\label{sec:ICARUS}

The Short Baseline Neutrino (SBN) Program~\cite{sbnproposal} primarily utilizes the Booster Neutrino Beam (BNB), which uses protons accelerated to 8\,GeV to produce a beam of neutrinos peaked between 500-600\,MeV at a distance of several hundred meters from the beam origin \cite{MicroBooNE:Flux}. 
With Short-Baseline Near Detector (SBND) located 110\,m from the beam origin and ICARUS at 600\,m, 
this program aims to clarify the sterile neutrino picture in light of reported anomalies, 
such as the LSND~\cite{LSND:Anomaly} and MiniBooNE~\cite{MiniBooNE:Anomaly} results.
In addition to BNB, neutrinos produced by the NuMI (Neutrinos at the Main Injector) beam at Fermilab \cite{NuMI:beam} also provide a significant flux of neutrinos at ICARUS, 800\,m away and 100\,mrad off-axis.
We describe the ICARUS detector and the NuMI beam in this section.

\subsection{The ICARUS detector}
\label{sec:ICARUSDetector}

Over the past few decades, liquid argon (LAr) time-projection chambers (TPCs) have emerged as a prominent medium for particle detectors, suitable for observing and measuring neutrino interactions. They operate primarily as a tracking calorimeter in which the charged particles resulting from neutrino interactions are observed. Charged particles ionize the argon medium as they propagate, and an applied electric field causes these electrons to drift toward sensing elements at the anode. Argon as a detector medium has the advantage of also scintillating due to the energy deposited by traversing charged particles, which provides a fast signal which can be used to provide timing information and a trigger~\cite{ICARUS:Trigger}.

The ICARUS (Imaging Cosmic and Rare Underground Signals) ``T600'' is a LAr TPC detector that was constructed and first operated in Italy~\cite{ICARUS:Construction}, sited at the Laboratori Nazionali del Gran Sasso (LNGS) measuring neutrinos produced at CERN~\cite{ICARUS:CNGS}. The ICARUS detector holds 760\,tons of LAr, of which 476\,tons constitutes active (instrumented) detector mass. After a period of refurbishment and upgrades at CERN, the ICARUS detector was shipped to Fermilab, near Chicago, to serve as part of the Short Baseline Neutrino (SBN) Program. It was installed in a shallow pit at Fermilab and covered with a 6\,m.w.e. overburden to reduce the low energy cosmic-ray background. 

The ICARUS detector is comprised of two modules or cryostats.
Each module contains two distinct volumes in a horizontal drift configuration, each 1.5\,m wide and sharing a single central cathode, with separate anodes located on the outer walls.
The electric field is held near 500\,V/cm during nominal operations. Each of these four TPC volumes has an active volume which is approximately 17.9\,m long and 3.2\,m tall. The sensing elements at the ICARUS anodes are three planes of wires, the first of which is in a horizontal orientation ($0^{\circ}$) and the second ($+60^{\circ}$) and third ($-60^{\circ}$) of which are angled. The wires are maintained with a bias voltage applied, causing the drifting electrons to flow past the first two planes, inducing signals on nearby wires, then collecting on the third plane. Combinations of wires with a signal yield information about the longitudinal and vertical dimension of charged particles and the time of signals yields the location in the drift dimension, thus providing a three-dimensional view of particles in the detector. 
Photomultiplier tubes (PMT) are used to detect scintillation light and check if activity is in-time with the beam. While the TPC would show activity in most 1.6 ms readout periods, only a fraction of spills will actually have activity coincident with the beam (which have spills of a few microseconds). Thus, the light is used to provide a trigger on activity in-time with the beam~\cite{ICARUS:Trigger} with a ns-level timing resolution.
90 PMTs are installed behind the wire plane of each TPC.
The deployment of ICARUS at Fermilab and initial operations have been described in Ref.~\cite{ICARUS:InstallFNAL}.

\subsection{The NuMI beam}
\label{sec:NuMIBeam}

As described in Sec.~\ref{sec:Intro}, this analysis uses the NuMI~\cite{NuMI:beam} beam as the neutrino source.
While the BNB uses 8\,GeV protons from the Booster impinged on a beryllium target, NuMI extracts protons from the Main Injector at 120\,GeV and collides them with a graphite target.
Downstream focusing horns redirect the resulting charged hadrons, 
which subsequently decay to produce neutrinos.
The system can selectively focus positively-charged hadrons to produce a predominantly neutrino beam (neutrino mode), or negatively-charged hadrons to produce an antineutrino beam (antineutrino mode).

\begin{figure}[htbp]
  \centering
  \includegraphics[width=0.45\textwidth]{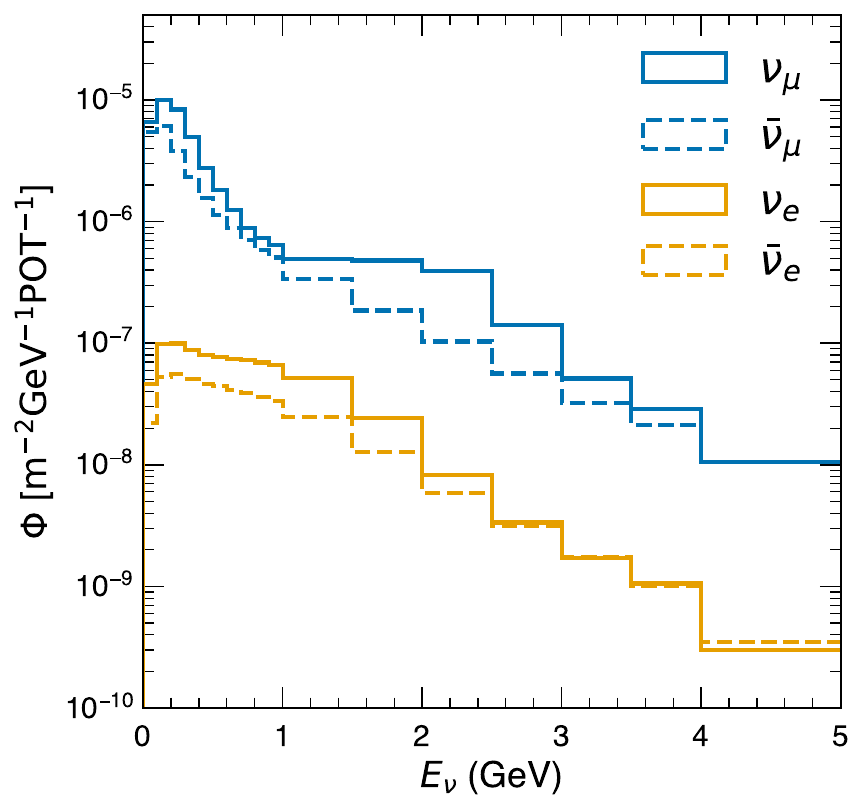}
  \caption{The simulated neutrino fluxes at the ICARUS detector.}
  \label{fig:ICARUSFlux}
\end{figure}

The configuration of targetry and focusing produces an on-axis flux of neutrinos peaked around 6\,GeV. 
However, at 100\,mrad off-axis, the kinematic properties of pion decay result in a neutrino flux shifted toward lower energies. Neutrinos produced by kaon decays at this off-axis angle create a pronounced shoulder in the flux near 2\,GeV.
A further consequence of the off-axis location is that a larger fraction of neutrinos reaching ICARUS originate from hadrons that were not forward-directed by the focusing system. 
Section~\ref{sec:eventsim:flux} comments on aspects of the NuMI flux simulation and the resulting spectra at ICARUS. Due to the shape of ICARUS and its location relative to the NuMI beam axis, the detector sweeps out a few mrad of off-axis angles around 100\,mrad (approximately 99 to 104\,mrad).
Figure~\ref{fig:ICARUSFlux} shows the simulated neutrino fluxes at the ICARUS detector from NuMI, which is described in Sec.~\ref{sec:eventsim:flux}.

Epochs of data-taking for the ICARUS experiment are defined based on the operational periods of the Fermilab accelerators and are labeled with a Run number numerically in increasing order. 
The data used in this analysis were collected during ICARUS Run 1 and Run 2, spanning from early June to early July 2022 and the end of February to late June 2023, respectively. 
During these periods, the NuMI beam operated in neutrino-mode.
This analysis utilizes a neutrino beam dataset corresponding to $2.5\times10^{20}$ protons on target (POT).

%% file: Sections/EventSimulation.tex
\section{Event simulation}
\label{sec:eventsim}

The simulation of neutrino events in the ICARUS detector is performed in several stages.
First, the NuMI neutrino flux is modeled to predict the kinematics of neutrinos reaching the detector.
Second, the interactions of these neutrinos with nuclei are simulated using a comprehensive cross-section model that accounts for nuclear effects and final-state interactions.
Third, the resulting particles, along with cosmogenic backgrounds, are propagated through a detailed model of the ICARUS detector geometry.
Finally, the induced electronic signals and scintillation light are simulated to produce raw data formats identical to those collected by the detector.
Each of these stages is described in detail in the following subsections.

\subsection{Neutrino flux simulation}
\label{sec:eventsim:flux}

The off-axis location of the detector results in a combination of low-energy ($0.5\,\mathrm{GeV}$) muon decays, moderate-energy ($0.5-2\,\mathrm{GeV}$) pion decays, and high-energy ($>2\,\mathrm{GeV}$) kaon decays in the neutrino flux reaching the detector.
The neutrinos produced in the NuMI beamline and their propagation to the ICARUS detector are simulated using flux predictions originally produced by the G4NuMI \GEANTfour package for NOvA oscillation analysis~\cite{PhysRevLett.123.151803}.
The Package to Predict the Flux (PPFX)~\cite{PPFX} is used to constrain the hadron production parameters from the external data.
PPFX provides a central value correction to the nominal flux simulation, along with the uncertainties described in detail in Sec.~\ref{sec:Systematics:Flux}.
Figure~\ref{fig:ICARUSFlux} shows the predicted neutrino fluxes at the ICARUS detector which are used to simulate neutrino interactions at the detector~\cite{wood2025numineutrinofluxprediction}.

\subsection{Neutrino interaction simulation}
\label{sec:eventsim:nugenerator}

The \GENIE event generator~\cite{Andreopoulos:2009rq,Andreopoulos:2015wxa} with \texttt{AR23\_20i\_00\_000} (``\texttt{AR23\_20i}") comprehensive model configuration (CMC) is used as the reference model to predict neutrino interactions.
The main components of the neutrino-nucleus interaction are the nuclear ground state, the neutrino-nucleon interaction, and the final-state interaction(s).
The initial state of the nucleon is based on the Local Fermi Gas (LFG) model, which relates the initial momentum to the removal energy.
The phase-space is further extended above the Fermi momentum, which corresponds to $\sim$20\% of nucleons inside the medium, measured from electron scattering data~\cite{RevModPhys.89.045002,PhysRevLett.106.052301}.

The quasi-elastic (QE) interaction is described by the Valencia model~\cite{PhysRevC.83.045501},
where the nucleon axial form factor is using the $z$-expansion formalism fit to neutrino-deuterium bubble chamber data~\cite{PhysRevD.93.113015} and the vector form factor is using the BBBA07 parameters~\cite{Bodek:2007ym}.
The interaction that involves multiple nucleons interacting with a neutrino, meson-exchange current (MEC), is simulated using the SuSAv2-MEC model~\cite{Gonzalez-Jimenez:2014eqa,Megias:2016fjk,RuizSimo:2016rtu,RuizSimo:2016ikw}.
The baryon resonance (RES) production channel is simulated using the Berger-Sehgal model~\cite{PhysRevD.76.113004,PhysRevD.79.079903}, which is based on the Rein-Sehgal model~\cite{REIN198179} but incorporates the necessary lepton mass effects.
A dipole ansatz is used to model the axial form factor of this channel with the value of mass $1.09\,\mathrm{GeV/c}$ and an overall cross-section scale factor of 0.84, which are extracted from a fitting to free nucleon datasets~\cite{GENIE:2021zuu}.
The resonance masses below $W_{\mathrm{cut}}=1.809\,\mathrm{GeV/c}$ are considered, which are also extracted from the same fitting procedure.
Deep inelastic scattering (DIS), which has hadronic mass above $W_{\mathrm{cut}}$, uses the Bodek-Yang model~\cite{ABodek_2003}, and is extrapolated down to the resonance regime to describe the shallow inelastic region (SIS).
The particles produced at the vertex may re-scatter or be absorbed before exiting the nucleus, and such a final-state interaction (FSI) is described by the hA2018 model~\cite{Dytman:2011zz,PhysRevD.104.053006}.
Neutral-current elastic scattering is simulated using the model described in Ref.~\cite{PhysRevD.35.785}.

\subsection{Simulation of cosmogenic particles}
\label{sec:eventsim:cosmicgenerator}

We use the \CORSIKA package~\cite{Heck:1998vt} to simulate showers initiated by cosmic ray particles, using proton-induced showers.
These simulated cosmogenic showers are added into the simulation chain, producing a combined set of generated particles from the \GENIE simulated neutrino interaction(s) and generated particles from \CORSIKA. This combined set of particles is then input to the detector simulation steps as described in Sec.~\ref{sec:eventsim:detsim}.
This procedure models the coincident cosmogenic background expected during a beam spill.
Cosmogenic activity inside the beam spill that causes a trigger is handled in a data-driven way as described in Sec.~\ref{sec:RecoSel:EvSel}.

\subsection{Detector simulation}
\label{sec:eventsim:detsim}

Final-state particles from \GENIE and \CORSIKA are propagated through the detector using \GEANTfour~\cite{Geant4:SimToolkit,Geant4:RecentDev}.
For each propagation step, \GEANTfour simulates energy deposition, which is used to produce a set of ionization electrons and scintillation photons.
LArSoft interfaces with these outputs and additional tool kits to simulate the drift of electrons toward the wire planes, accounting for diffusion, electron lifetime (due to absorption), and recombination effects.

The simulation of the induction and collection signals at the wires is performed using the Wire-Cell Toolkit~\cite{MicroBooNE:SignalProc}.
This toolkit models the detector response by calculating the current induced on the wires as electrons drift through a two-dimensional (time and wire) domain.
Customized response functions and calibrations are applied to match the specific characteristics of the ICARUS detector~\cite{ICARUS:Calibration}.
This process results in simulated waveforms for each wire, which are then processed using the signal processing and reconstruction chain detailed in Sec.~\ref{sec:RecoSel:Reco}.
The angular-dependence of ionization recombination~\cite{ICARUS:AngRecomb} was not included in the simulation, and was corrected in data via the calorimetry.

Scintillation is also simulated in the detector and feeds into a simulation of the trigger logic based on the signals at the photomultiplier tubes. 
As the simulated trigger was generally found to have a higher efficiency than that found for data in Ref.~\cite{ICARUS:Trigger}, weights are evaluated by taking the data-to-simulation ratio of the trigger efficiencies as a function of visible energy in each spill, and are applied to the simulation.




%% file: Sections/EventRecoAndSelection.tex
\section{Event reconstruction and selection}
\label{sec:RecoSel}

This section details the event reconstruction chain used to identify the signal events, followed by the definition of CCQE-like signal topology. 
A series of event selection criteria is applied to isolate signal candidates while suppressing backgrounds.
Finally, we describe the implementation of sideband samples, which are  used to constrain background events containing pions in the final state.

\subsection{ICARUS reconstruction}
\label{sec:RecoSel:Reco}

The primary detector subsystem used is the charge readout from the TPC. The LAr TPC readout waveforms collected from an individual readout window (an approximately 1.6\,ms window for each 9.5\,$\mu$s beam spill) from data and simulation undergo several steps of signal processing and reconstruction.

The effects of the field and electronics response are deconvolved from the raw waveforms. 
A one-dimensional deconvolution is performed in the time domain to account for the electronics response and the time-dependent current induced on the wires by passing ionization electrons.

The resulting processed signals are proportional to the number of ionization electrons arriving at each wire, prior to corrections for high-level detector effects such as electron recombination or attenuation due to the electron lifetime.
To improve the signal-to-noise ratio, a coherent noise subtraction is performed to mitigate correlated noise across multiple wire channels.

The waveforms are then reduced to sparse regions of interest (ROIs) where signal is identified.
These ROIs are fitted with one or more Gaussian functions to represent the signal as a set of ``hits," each characterized by a peak time, width, and charge. These hits are filtered by an algorithm which rejects hits on a given plane with no corresponding spatial match on other planes as likely noise.

For event-level reconstruction, we use the Pandora~\cite{PandoraPFA} multi-algorithm pattern-recognition software available in LArSoft and used in numerous LAr TPCs~\cite{MicroBooNE:Pandora,ProtoDUNE:Pandora}. 
The filtered set of hits serve as the input to Pandora, and a separate instance is run on the hits from each cryostat.
Within the readout window for a given spill, Pandora returns a set of reconstructed interactions that are either tagged as ``unambiguously cosmic'' or treated as a ``neutrino candidate."
The reconstruction returns a suite of information, which includes a reconstructed vertex location and a set of reconstructed particles with a hierarchy.
For each reconstructed particle additional information is available such as its characterization as track-like or shower-like, the hits in the detector that went into making the reconstructed particle, and the reconstructed three-dimensional positions of points from projection matching. 

The reconstructed particles from candidate neutrino interactions then undergo additional characterization under a track-like hypothesis and shower-like hypothesis. Each is described separately below. 
For the track hypothesis, these steps include trajectory fitting, calorimetry to determine differential energy deposition per unit length (dE/dx) along the trajectory for each plane and particle identification based on comparing the dE/dx versus residual range for each plane to the expected curves for different particle types (proton, muon, charged pion, and kaon).
The beginning and ending points of the track trajectory provide the start and stop points and directions of the reconstructed particle. 
Momentum of contained track-like particles is determined by application of the Continuous Slowing Down Approximation (CSDA)~\cite{PDG}.
For the shower hypothesis, these steps include calorimetry, dE/dx in the initial portion, determining the gap between the vertex and the start of the shower, etc. 
A Boosted Decision Tree (BDT) algorithm assigns a ``track score" to each reconstructed particle, which is used to classify objects as either tracks or showers. 

\subsection{Signal definition}
\label{sec:RecoSel:SigDef}

Our goal is to isolate a sample of CCQE-like interactions of neutrinos on argon in ICARUS.
The signal events are defined as interactions in which a muon neutrino or antineutrino undergoes a CC interaction producing at least one proton, any number of neutrons, and no other charged or neutral hadrons in the final state. 
Additional phase space constraints are applied in the definition of the signal.
Muons are required to have momentum above 226\,MeV/c, corresponding to a distance of approximately 50\,cm traveled in liquid argon.
The leading proton in the final state is required to have momentum between 400 and 1000\,MeV/c. 
This requirement excludes regions where the proton selection efficiency drops below $\sim25\%$. At low momenta, efficiency is restricted by track length, whereas at high momenta, protons frequently rescatter before they can be identified via their characteristic dE/dx profile.
Additional protons in the final state are allowed, with no lower bound placed on their momenta. 
Any number of neutrons in the final state is allowed.
We require no pions or other mesons, no heavy excited baryons in the final state along with no photons above 10\,MeV~\cite{MINERvA:2019Ruterbories}.

\begin{figure*}
    \centering
    \includegraphics[width=0.90\textwidth]{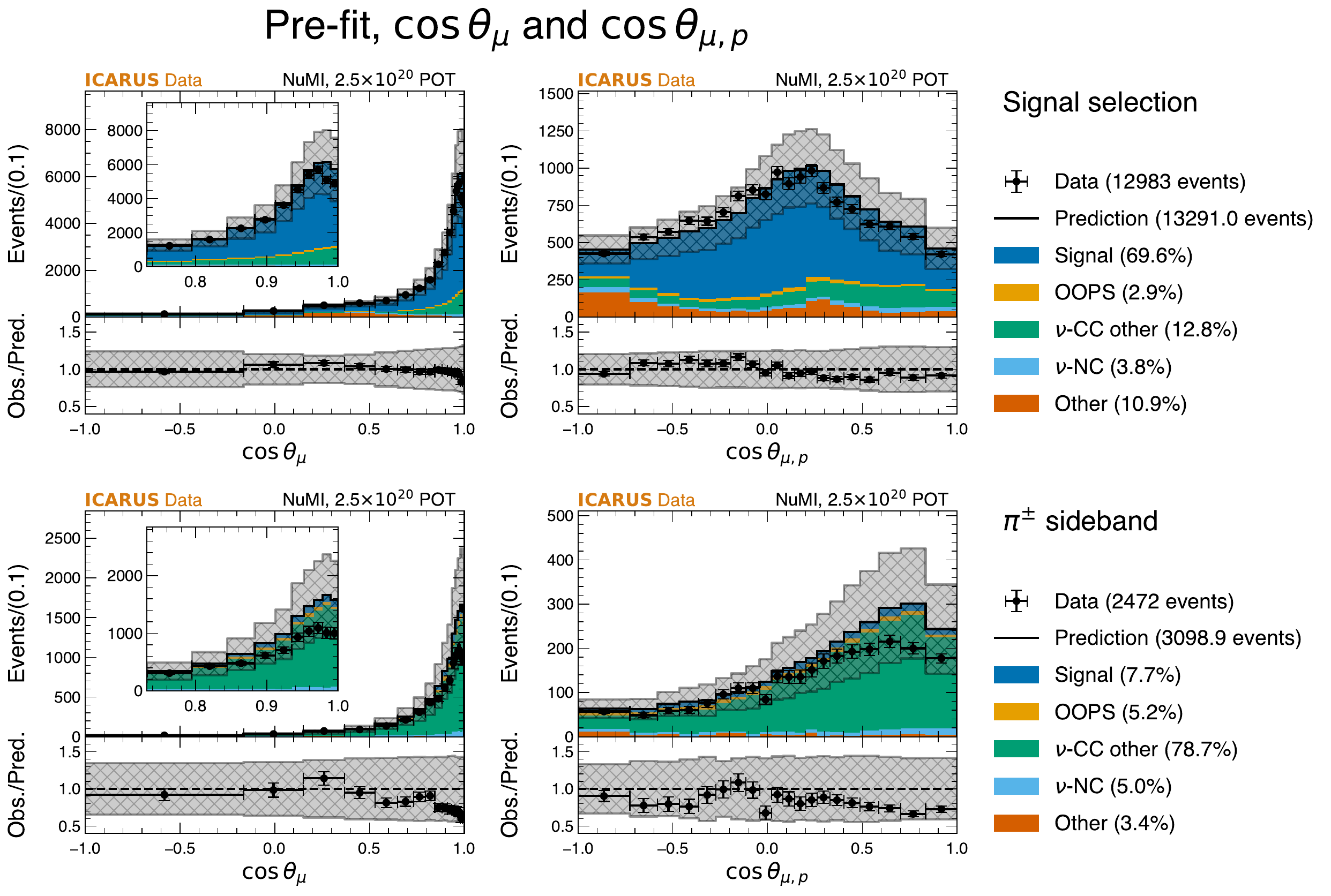}
    \caption{Reconstructed $\cos{\theta_\mu}$ (left) and $\cos{\theta_{\mu,p}}$ (right) distributions before the fit (``pre-fit"). Events that pass signal (sideband) selection described in Sec.~\ref{sec:RecoSel:EvSel} (\ref{sec:RecoSel:Sideband}) are shown in the top (bottom) plots. The gray hatched bands represent the statistical and systematic uncertainties on the prediction.}
    \label{fig:PrefitHist_Angular}
\end{figure*}

\begin{figure*}
    \centering
    \includegraphics[width=0.90\textwidth]{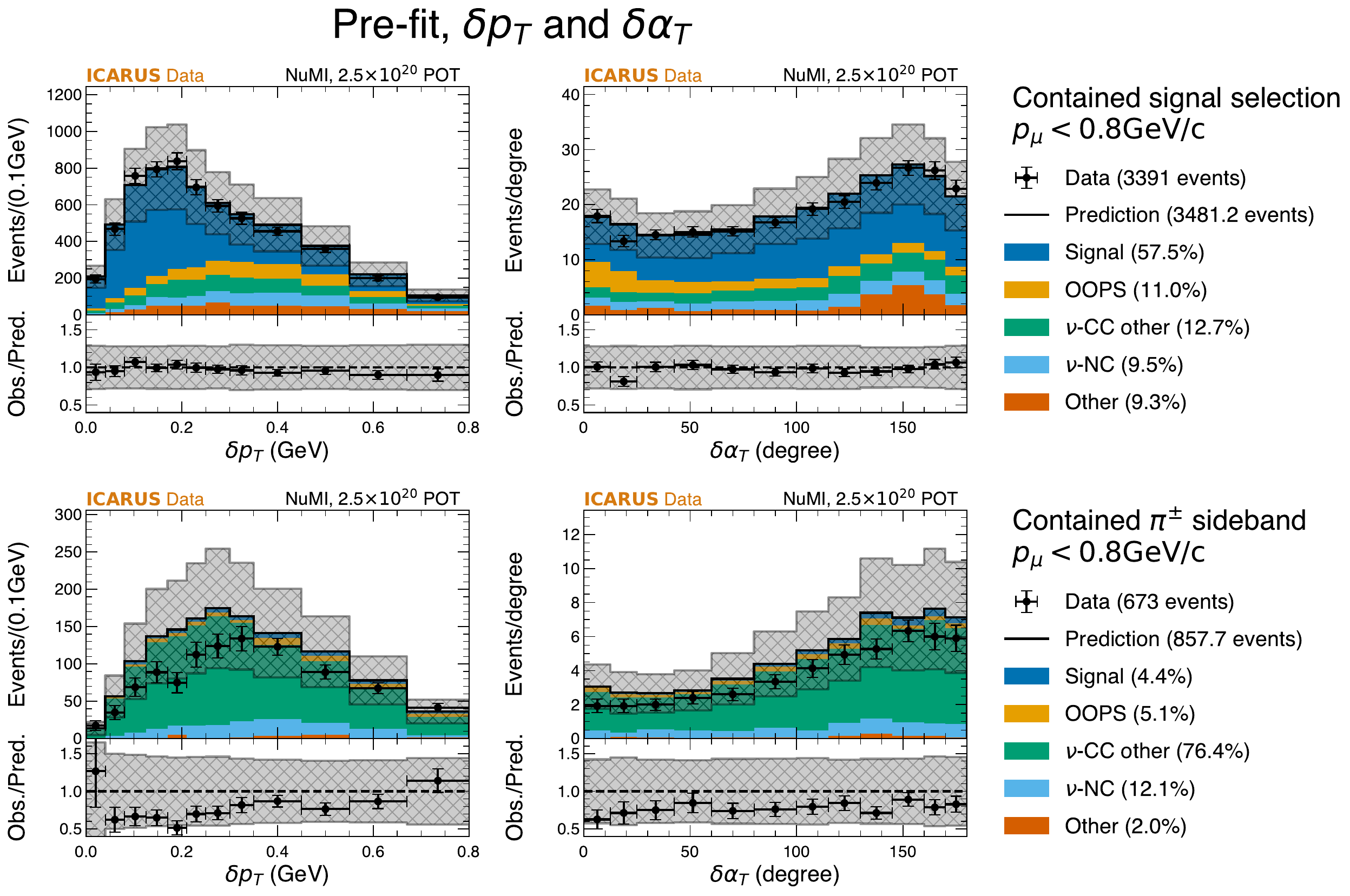}
    \caption{Reconstructed $\delta p_{T}$ (left) and $\delta \alpha_{T}$ (right) distributions before the fit (``pre-fit"). Events that pass signal (sideband) selection described in Sec.~\ref{sec:RecoSel:EvSel} (\ref{sec:RecoSel:Sideband}) are shown in the top (bottom) plots. The gray hatched bands represent the statistical and systematic uncertainties on the prediction.}
    \label{fig:PrefitHist_TKI}
\end{figure*}

\subsection{Event selection}
\label{sec:RecoSel:EvSel}

We select neutrino events originating in the detector fiducial volume with final-state muon and proton candidates with no other final-state particle. 
Data quality criteria are applied based on beam and detector operations to ensure that spills are close enough to nominal expectations. 
Beam quality cuts are similar to cuts used by NOvA~\cite{NOvA:FirstResults}. 
Data quality considerations were largely applied on entire DAQ operation sessions (referred to as ``runs,'' which may last for only a short period of time or for multiple days). 
A set of good runs feed into this analysis as determined from operating and trigger conditions and by checking that high-level reconstruction metrics are not anomalous for that run. A Kolmogorov-Smirnov test was used to compare the fractional accumulation of POT against the candidates for both the full and contained-only selections across Runs 1 and 2. All tests yielded $p$-values greater than 0.05, confirming consistency.
For simulated events, an additional cut is added requiring the spill to have an emulated trigger.
Events tagged as unambiguous cosmic are excluded from the selected sample.
The reconstructed vertex is required to be in the fiducial volume. 
The fiducial volume is defined by applying offsets from the active volume boundaries of each cryostat: 25\,cm in the drift and vertical directions, 30(50)\,cm from the upstream (downstream) in the beam direction.
Two regions are further fiducialized to mitigate known detector defects which correspond to 1.9\% of the volume described above.
In total, this yields a fiducial volume of $2.21\times10^{8}\,\text{cm}^{3}$, which is 65.7\% of the active volume.
The mass of the fiducial volume is 306.9\,t.

Candidate muon selection begins by using a BDT which assigns a track score to each particle.
To match the signal definition, the event selection requires a candidate muon track that is tagged as a primary particle by Pandora, starts within 10\,cm of the reconstructed vertex, and has a length exceeding 50\,cm. 
A track is defined as ``contained" if its endpoint is located more than 10\,cm away from the active volume boundary.
For contained muon candidates, the particle identification (PID) scores must be more consistent with a muon than a proton. If the track exits the detector, the PID score is not checked, as the Bragg peak will not be seen at the end of the muon trajectory. The longest track satisfying these requirements is selected as the muon candidate.

Proton candidates are selected from the remaining tracks by requiring them to be tagged as primary particles starting within 10\,cm of the vertex. These tracks must be contained and have PID scores more consistent with a proton than a muon. The event is required to contain at least one such proton candidate, with the leading proton required to have a reconstructed momentum between 400\,MeV/c and 1000\,MeV/c.

To reduce backgrounds, additional selection criteria are applied to the hadronic system to reject events containing pions in the final state. 
First, all primary tracks starting within 10\,cm of the vertex (excluding the muon candidate) must be contained.
To reject charged pions, events containing a second track that satisfies muon identification criteria are rejected.
Finally, neutral pion backgrounds are mitigated by identifying photon-induced showers.
Photon candidates are defined as shower objects with a length exceeding 10\,cm and a starting point displaced by at least 5\,cm from the primary vertex. Any event containing such a photon candidate is rejected.

An additional containment requirement on the muon is applied to construct the TKI variables which require precise momenta of both the muon and proton. The track momenta are determined via CSDA based on track range, which has better resolution than the multiple Coulomb scattering (MCS) method~\cite{MCS}. 
Requiring muon containment rejects events where the muon momentum is above about 1\,GeV/c. 
A further phase-space restriction of $p_{\mu}<0.8$\,GeV/c is added to both the signal definition and the event selection for TKI variables.

The top plots of Figure~\ref{fig:PrefitHist_Angular} and \ref{fig:PrefitHist_TKI} show the signal distribution for the kinematic variables of interest. 
The predicted event rates are subdivided into categories as discussed below.
The ``Signal" is the contribution coming from events that satisfy the signal definition.
Events that are signal-like but fail muon or proton momentum requirements fall into the out-of-phase-space category (``OOPS").
The CC events that are not classified as ``Signal" or ``OOPS" are categorized as ``$\nu$-CC other", while neutral-current interactions are placed in the ``$\nu$-NC" category.
The ``Other'' category consists of events not matched to neutrino interactions and is dominated by two types of cosmogenic activity. 
The first consists of events triggered by a neutrino interaction with coincident cosmogenic activity occurring within the same TPC readout window; the \CORSIKA package~\cite{Heck:1998vt} is used to simulate this contribution.
The second type consists of events triggered by cosmogenic activity during the beam spill.
We utilize a data-driven approach using a sample of triggered data collected outside of beam spill times. 
This sample is reweighted to match the exposure of the beam data and contributes approximately half of the ``Other" category.

\subsection{Sideband for pion events}
\label{sec:RecoSel:Sideband}

\begin{figure}[htbp]
  \centering
  \includegraphics[width=0.23\textwidth]{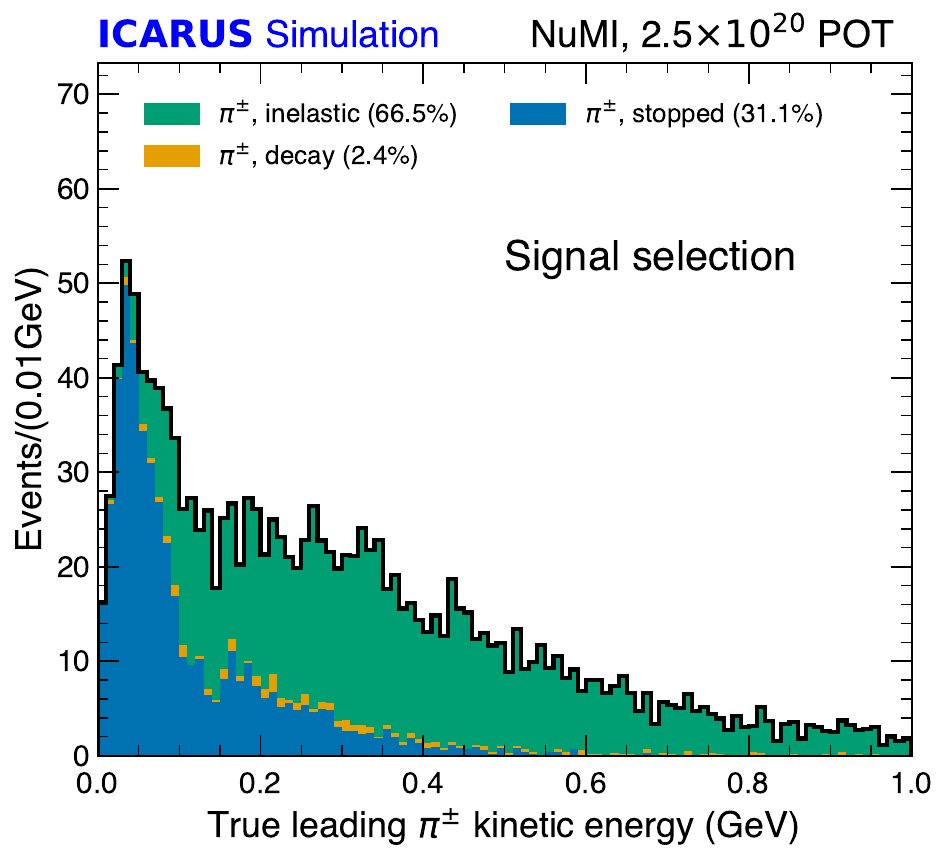}
  \includegraphics[width=0.23\textwidth]{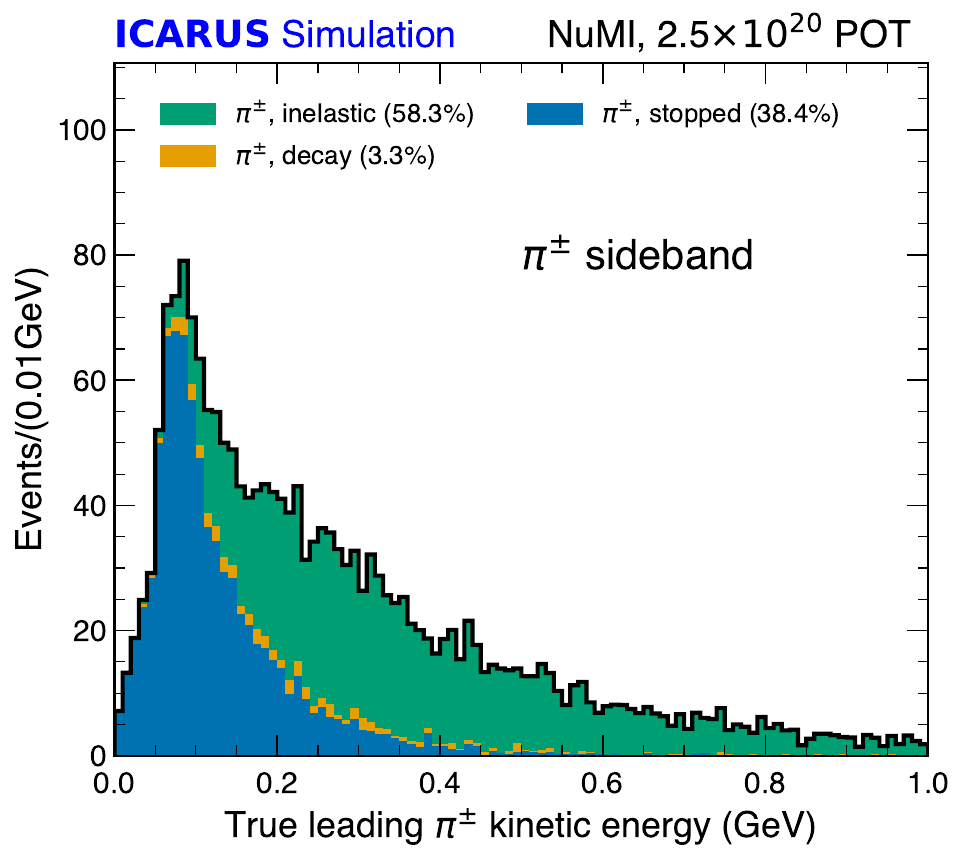}
  \caption{The true leading pion kinetic energy for the interactions passing signal (left) and sideband (right) selection.}
  \label{fig:SidebandTruePionKE}
\end{figure}

Neutrino interactions with charged pions in the final state can pass the signal selection if the pion is not identified.
A sideband is defined by inverting the charged pion-veto condition, while keeping all other requirements in the signal selection. 
That is, we require the presence of a ``tagged'' charged pion candidate.
We further require such a candidate to be at least 10\,cm long to define a higher purity sample of charged pion-producing interactions for the sideband, and to avoid signal protons mis-identified as pions.
Figure~\ref{fig:SidebandTruePionKE} shows the kinetic energy of the true leading charged pion for the neutrino interactions passing signal and sideband selection.
The background from charged pions with length less than 10cm in the signal sample, which are not included in this sideband, is non-negligible.

\begin{figure}[htbp]
    \centering
    \includegraphics[width=0.23\textwidth]{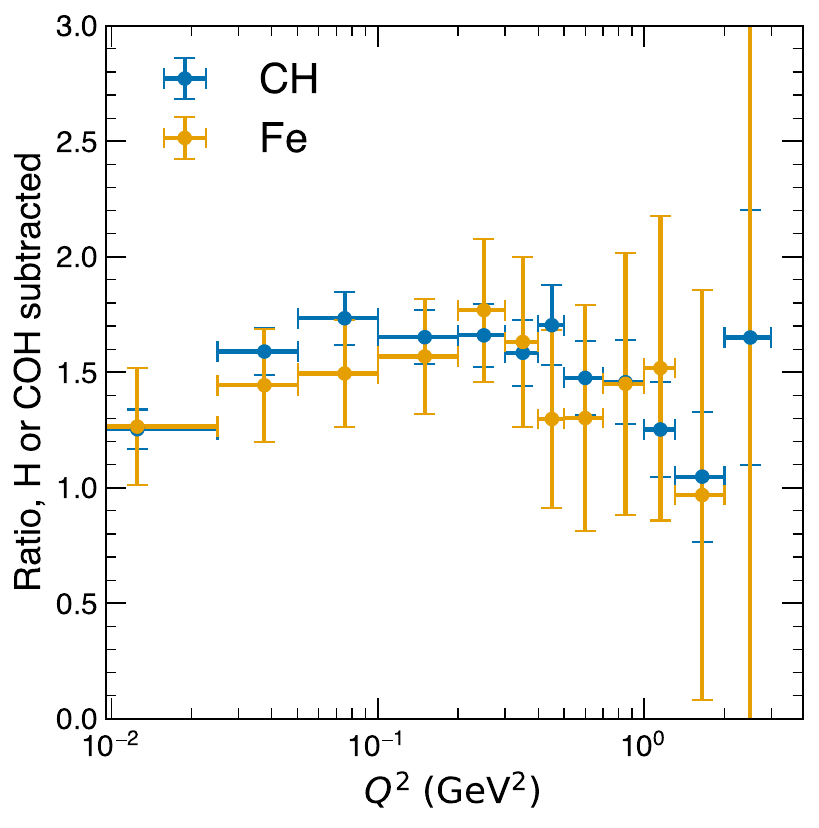}
    \includegraphics[width=0.24\textwidth]{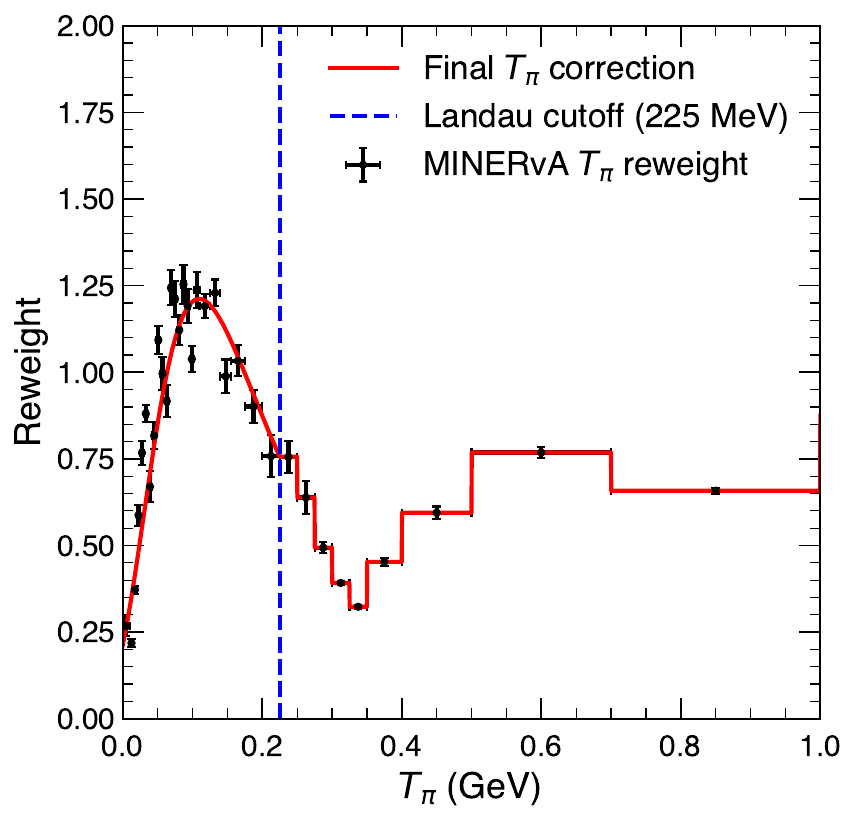}
    \caption{(Left) The data-to-prediction ratio of differential cross-section binned in $Q^{2}$ from the MINERvA measurement~\cite{PhysRevLett.131.011801}, where the contribution of hydrogen or coherent pion production are subtracted. (Right) The scale factor as a function of true pion kinetic energy ($T_{\pi}$) extracted to describe the discrepancy in pion range from Ref.~\cite{mehreenthesis}.}
    \label{fig:SPPCVCorrection}
\end{figure}

The measurement of the single pion production cross section from MINERvA~\cite{PhysRevLett.131.011801} is used to extract a correction and uncertainty on their contribution.
The measured differential cross section binned in $Q^{2}=-q^{2}$, where $q$ is the four-momentum transfer from the neutrino, is compared to the prediction from the reference model.
The difference between the data and the model after subtracting the contribution from hydrogen and coherent pion production (COH) is corrected using a bin-by-bin scale factor.
The $Q^{2}$-binned single pion production scale factor was extracted from scintillator data (CH) and showed reasonable agreement to the same value extracted from the iron (Fe) measurement.
The residual difference in the pion kinematics was also examined, but due to the threshold of $T_{\pi}>35\,\mathrm{MeV}$ in Ref.~\cite{PhysRevLett.131.011801}, 
we used the measurement from Ref.~\cite{mehreenthesis}, where the pion kinetic energy goes down to zero.
The discrepancy in the pion candidate range is corrected by pion kinetic energy, $T_{\pi}$-binned scale factor (black marker in the right plot of Fig.~\ref{fig:SPPCVCorrection}).
The measured correction value is fit with a Landau function for $T_{\pi}<225\,\mathrm{MeV}$.
The single-pion production events are weighted by $Q^{2}$ and $T_{\pi}$ scale factors in Fig.~\ref{fig:SPPCVCorrection}.
The central value predictions of the sideband are shown in the bottom plots of Fig.~\ref{fig:PrefitHist_Angular} and \ref{fig:PrefitHist_TKI}.

%% file: Sections/Systematics.tex
\section{Systematic uncertainty}
\label{sec:Systematics}

The systematic uncertainties associated with the predicted event rates and kinematic distributions are categorized into four primary sources: the NuMI neutrino flux, the detector response, the neutrino-nucleus interaction model, and the modeling of particle propagation.
The total uncertainty is dominated by the detector response, which has the largest impact on the predicted event normalization.
While uncertainties related to the interaction modeling of signal events largely cancel out through the efficiency calculation,
the interaction modeling for background processes remains the second most significant source of systematic uncertainty.
An overview of these sources is provided in this Section, with a detailed breakdown of their impacts discussed in the following subsections.

\subsection{Flux uncertainties}
\label{sec:Systematics:Flux}

The nominal simulated prediction is reweighted to G4NuMI (v24.10.0) \GEANTfour (v4.10.4) simulated flux along with 
NA49~\cite{NA49PionFrompC,Tinti:2010zz,thena49collaboration2013inclusiveproductionprotonsantiprotons} (p+C collision data with proton momentum, $p_p = 158$\,GeV/c) and 
MIPP~\cite{PhysRevD.90.032001} (proton on NuMI replica target with $p_p = 120$\,GeV/c) hadron production data using PPFX~\cite{PPFX} (v02.11.05). 
An uncertainty band is generated by propagating the errors on the data along with conservative uncertainty estimates when data is scaled (to other neutrino energy and nucleus) 
and for processes outside the data regions altogether. 
Modifications were made to the baseline PPFX code to account for effects that are important at the ICAURS off-axis angle relative to the near-on-axis beam sampled by NuMI-based experiments; 
specifically the higher rates of proton interactions with the aluminum horns and steel shielding, as well as the higher proportion of meson reinteractions, 
in the interaction/decay chains that produce neutrinos directed toward the ICARUS detector. 
New degrees of freedom were introduced into the PPFX error propagation to account for these differences, 
including new regions of phase space and additional nuclear targets. 
The large uncertainty (40\%) ascribed to unmeasured, very low four-momentum transfer quasi-elastic p-C collisions, which leave the kinematics of the outgoing proton essentially unchanged, were removed.
A bug that impacted the calculation of meson reinteraction kinematics was also fixed~\cite{wood2025numineutrinofluxprediction}.

The uncertainties originating from the focusing of the NuMI beam are evaluated from statistically independent flux predictions with altered parameters:
\begin{itemize}
    \item Horn current $200 \pm 2\,\mathrm{kA}$
    \item Horn 1 position shifted $\pm 3\,\mathrm{mm}$ in both directions perpendicular to the beam direction
    \item Beam width $1.5 \pm 0.2\,\mathrm{mm}$
    \item Horn 2 position shifted $\pm 3\,\mathrm{mm}$ in both directions perpendicular to the beam direction
    \item Horn water layer $1 \pm 1\,\mathrm{mm}$
    \item Beam position shifted $\pm 1\,\mathrm{mm}$ in both directions perpendicular to the beam direction
    \item Target position $\pm 7\,\mathrm{mm}$ along the beam direction
\end{itemize}

The uncertainty on the integral of muon and anti-muon fluxes for the neutrino energy between $0$ and $20\,\mathrm{GeV}$ is $11.0\%$ from hadron production and $1.7\%$ from beam focusing (updated internal numbers based on Ref.~\cite{wood2025numineutrinofluxprediction}).

\subsection{Detector uncertainties}
\label{sec:Systematics:Detector}

The majority of the detector uncertainties are assessed by re-simulating samples with altered detector parameters and comparing them to the central value (CV) simulation, following the methodology in Ref.~\cite{PhysRevLett.134.151801}.
To mitigate the impact of limited sample statistics, a common set of signal-like neutrino interactions without cosmic backgrounds are used as the input.

As described in Sec.~\ref{sec:eventsim:detsim}, the detector parameters for the CV simulation are calibrated following the procedures in Ref.~\cite{ICARUS:Calibration}.
The remaining channel-to-channel variations observed in low-level variables, such as the charge per unit length (dQ/dx), are treated with detector uncertainties.
The gain and noise levels for the first wire plane were varied by $\pm20\%$ and $\pm10\%$, respectively, to encompass the variation of the peak position and width in the dQ/dx distribution for each channel from the cosmic muon samples.
The resulting deviations from the CV were evaluated in the reconstructed proton momentum distribution and found to be well-covered by flat normalization uncertainties of 3.5\% and 10.5\%, respectively.
The signal shape tuning procedure was found to be less reliable in describing the dQ/dx distribution on the first wire plane than the other two planes. 
Unlike in Ref.~\cite{ICARUS:Calibration}, the tuning in the first wire plane is still applied to the CV simulation, as it was also in Ref.~\cite{PhysRevLett.134.151801}. Similar to that publication, an additional uncertainty is considered for the first wire plane by inspecting the difference in the proton momentum between the CV and the un-tuned simulation: this variation was found to be dependent on proton momentum, varying from approximately 15\% at low momenta to 5\% at higher momenta.
The channel-to-channel transparency variations of the second wire plane are accounted for by a 7.9\% normalization uncertainty.
The variations in calorimetry (dE/dx) and electron lifetime were found to contribute 3.0\% and 5.9\%, respectively.

A data-driven approach was developed to evaluate the residual uncertainty associated with proton identification efficiency.
The method first collects high-purity proton candidate tracks from both data and simulation, identified by their characteristic ionization profiles and PIDA scores~\cite{larsoft:chi2pidalg}.
The initial portion of these tracks are systematically removed while preserving the end of the trajectory, and thus the Bragg peak, to emulate a lower-energy proton.
The reconstruction is re-run on these shortened tracks to produce proton tagging efficiency as a function of track range.
When scaled by a flat 11.8\%, the tagging efficiency in simulation was found to be in reasonable agreement with an envelope produced from shortened proton candidates in systematically shifted samples. This is captured in the analysis as an 11.8\% flat normalization uncertainty on proton identification.

The tracks were found to split more frequently in data than simulation near the cathode and the wire support structure, located in the middle of the detector length-wise, due to unsimulated geometric effects and imperfections.
The tracks that cross these boundaries from the simulation are systematically split into two tracks, based on the probability to be split as measured from the data used in Ref.~\cite{icaruscollaboration2026searchsterileneutrinooscillation}.
A 10\% uncertainty is assigned to the migration between the contained and exiting muon samples to account for the impact of this effect on the selection.

\subsection{Neutrino-nucleus interaction uncertainties}
\label{sec:Systematics:Interaction}

The variation of physics parameters are accessed through a reweighting on the centrally produced simulation events. The axial form factor of QE interactions are varied by the uncertainty obtained in the deuterium fit~\cite{PhysRevD.93.113015}.
The correlation between the four free parameters of the $z$-expansion formula are considered through a principal component analysis (PCA) of the error matrix.
The strength of the low-$Q^{2}$ (RPA correction~\cite{PhysRevC.83.045501}) and Coulomb correction are also varied. 
A reweight of the QE vector form factor to the dipole ansatz from BBBA07~\cite{Bodek:2007ym} is considered as an uncertainty.
The normalization for both charged- and neutral-current MEC events are varied by $\pm50\%$ as one standard deviation.
The two nucleons from the cluster of the MEC events are distributed isotropically in the cluster rest frame.
An alternative model with an angular distribution of $(3/2)\cos^{2}{\theta_{N}}$ where $\theta_{N}$ is the angle between the nucleons along the momentum transfer direction in the cluster rest frame is considered as an uncertainty.
The axial and vector form factor of the RES events are varied through masses in the dipole assumption.
The branching fraction of resonance decays with a photon ($\gamma$) or $\eta$-meson in the final states are varied by 50\%.
The decay angle of $\Delta \rightarrow N + \pi$ ($\Delta \rightarrow N + \gamma$) is varied to isotropic ($(3/2)\cos^{2}{\theta_{N}}$) and considered as an uncertainty.
As described in Sec.~\ref{sec:RecoSel:Sideband}, events with single pion in the final states are weighted by the scale factor extracted using MINERvA data.
Uncertainties are assigned for each $Q^{2}$ and $T_{\pi}$ correction separately where the size of one standard deviation is defined by the amount of each correction.
With the single pion production correction applied, we observed residual discrepancies in the sideband with forward-going muons; however, there is a large spread of current models describing pion productions~\cite{MIINERvASIS}.
For the charged-current interactions that are not signal (``$\nu$-CC other" category), additional template parameters are added to the fit (where the framework is described in Appendix~\ref{sec:appd_gundam_xsecextraction}) with prior uncertainties inspired by a comparison between generators:
\begin{itemize}
    \item Binning for the ``$\nu$-CC other," $(W,Q^{2})$-binned template parameters:
    \begin{itemize}
        \item $W$: below or above 1.4\,GeV
        \item $Q^{2}$: uniform binning between 0 and $0.8~\text{GeV}^{2}$, bin size of $0.1\,\text{GeV}^{2}$
    \end{itemize}
    \item Prior uncertainty of 40 (60)\% for $W$ below (above) 1.4\,GeV.
\end{itemize}
Scale factors for non-resonance events are varied by $\pm50\%$  per neutrino flavor, per nucleon type, per charged- and neutral-current, and per hadron multiplicity.
For the DIS events, the two ``higher twist" parameters, introduced in Ref.~\cite{ABodek_2003} that modifies the Nachtmann variable ($\xi$)~\cite{NACHTMANN1973237} to capture non-perturbative effects at low $Q^{2}$ and high Bjorken $x$, are varied.
The two parameters correcting the parton distribution function (PDF) of $u$ valence quarks in leading order GRV98 PDF set~\cite{Gluck:1998xa} are varied.
The uncertainty of the FSI are obtained by altering the mean-free-path and rescattering probability inside the nucleus, separately for the nucleon and pion.
The normalization and shape of the axial form factor of neutral- current elastic scattering events are varied.

\subsection{Particle propagation uncertainties}
\label{sec:Systematics:Propagation}

As the analysis is placing selection cuts to enhance signal or sideband events by accepting or rejecting particles tagged as protons or charged pions, the fates of these produced particles in the ICARUS detector can alter what is selectable or not. For example, if a proton undergoes an inelastic scatter and results in the production of other particles, the proton may not reach rest via ionization and be missed due to lack of a Bragg peak. Likewise, if a pion scatters, it may be missed and selected as a background event in the signal for example. Therefore, we turn to the \GEANTfourRW package~\cite{Calcutt_2021}, which provides a mechanism to reweight events based on, for example, the proton and pion contents of the events and the fates they encountered in the detector simulation multiplied across the steps the particles took. For this analysis, the \GEANTfourRW package was not fully working in the SBN software, but we were able to bring in a version built by the MicroBooNE collaboration (v01\_00\_03e). 
The simulation then injects weights for protons and pions with 1000 universes simulated, from which covariance matrices were evaluated based on the number and energies of protons and pions separately, using the simulated events that pass signal or sideband selection.
These covariance matrices were passed to the fitting framework (discussed in Appendix~\ref{sec:appd_gundam_xsecextraction}) directly for use in the analysis.

\subsection{Summary of systematic uncertainties}

The systematics uncertainties described above are incorporated into the fitting and extraction procedures detailed in Sec.~\ref{sec:extraction}.
As the analysis employs a simultaneous fit of the signal and sideband selections rather than a background subtraction,
the uncertainties on the event rates do not translate directly to the uncertainties on the extracted cross section due to the resulting correlation between the parameters.
To quantify the individual contributions to the final result, in Sec.~\ref{sec:fracuncxsec}, we used a method removing one category of systematic uncertainty at a time from an Asimov fit to illustrate the impact of each on the measured cross section.

%% file: Sections/CrossSectionExtraction.tex
\section{Cross-section extraction}
\label{sec:extraction}

We use the \GUNDAM fit and cross-section extraction package~\cite{GUNDAMRepo} developed initially within T2K and made available for broader use.
In Appendix~\ref{sec:appd_gundam_xsecextraction}, we illustrate how the unfolding, efficiency correction, and cross-section extraction is performed by \GUNDAM with an example.
In this section, we present the results of fit and extraction using the events and uncertainties described through Sec.~\ref{sec:RecoSel}--\ref{sec:Systematics}.
The fractional contribution on the cross-section from each source of uncertainty described in Sec.~\ref{sec:Systematics} is also shown.

\subsection{Fit}
\label{sec:xsecextraction:fitting}

As described in Appendix~\ref{sec:appd_gundam_xsecextraction}, a fit is performed prior to the cross-section extraction stage.
The goal of the fit is to find the set of parameters with the best agreement to the data, in the distributions of reconstructed variables.
The signal events are scaled freely (up to the regularization that is described at the end of this section) through the ``template parameters" which are defined for each cross-section bin.

The signal (Sec.~\ref{sec:RecoSel:EvSel}) and sideband (Sec.~\ref{sec:RecoSel:Sideband}) are fit simultaneously which provides a data-driven background constraint.
The two angular variables, $\cos{\theta_\mu}$ and $\cos{\theta_{\mu,p}}$, and the two TKI variables, $\delta p_{T}$ and $\delta \alpha_{T}$, are fit simultaneously as they are using the same selected events.
The statistical correlation between the two variables are considered through the method described in Ref.~\cite{gardiner2024mathematicalmethodsneutrinocrosssection}.
A covariance matrix, $C_{\mathrm{Stat.}}$, is constructed combining the bins of the signal and sideband histogram from both variables.
The diagonal components of  $C_{\mathrm{Stat.}}$ represent the usual bin-by-bin statistical uncertainty, whereas the off-diagonal components take into account the correlation.
The statistical part of the total log-likelihood (LLH) is calculated as 
\begin{align}
\label{eq:LLHStatCov}
\mathrm{LLH}_{\mathrm{Stat.}} = (\vec{D} - \vec{M})^{T}\boldsymbol{C}^{-1}_{\mathrm{Stat.}}(\vec{D} - \vec{M}),
\end{align}
where $M$ and $D$ represent predicted and observed number of events for each bin in the reconstructed space, respectively.
In addition to $\mathrm{LLH}_{\mathrm{Stat.}}$,
the systematic uncertainties are considered via the penalty term:
\begin{align}
\label{eq:LLHSyst}
\mathrm{LLH}_{\mathrm{Penalty}} = (\vec{\theta} - \vec{\theta}_{\mathrm{Prior}})^{T}\boldsymbol{C}^{-1}_{\mathrm{Penalty}}(\vec{\theta} - \vec{\theta}_{\mathrm{Prior}})
\end{align},
where $\vec{\theta}$ ($\vec{\theta}_{\mathrm{Prior}}$) is a vector of (prior) nuisance parameters assigned for each systematic uncertainty source and $\boldsymbol{C}_{\mathrm{Penalty}}$ is the covariance matrix between them.
The signal template parameters are assigned for both of the variables that are fit simultaneously.
Each signal event is then scaled by the two sets of template parameters (one for each variable), hence a strong anti-correlation is expected between the two sets.
A multiplication of two Gaussian responses, like the two template parameters, is non-Gaussian when the standard deviations are small enough~\cite{James:2006zz}.
We use a log-normal distribution to describe the template parameters. The parameter of interest ($\theta_\text{Template}$) follows Gaussian distributions, and their response is $e^{\theta_\text{Template}}$.
The $(W,Q^{2})$-binned template parameters that are applied to ``$\nu$-CC other" events (Sec.~\ref{sec:Systematics:Interaction}) are also treated with log-normal uncertainties due to their large prior uncertainty.

\begin{figure*}
    \centering
    \includegraphics[width=0.90\textwidth]{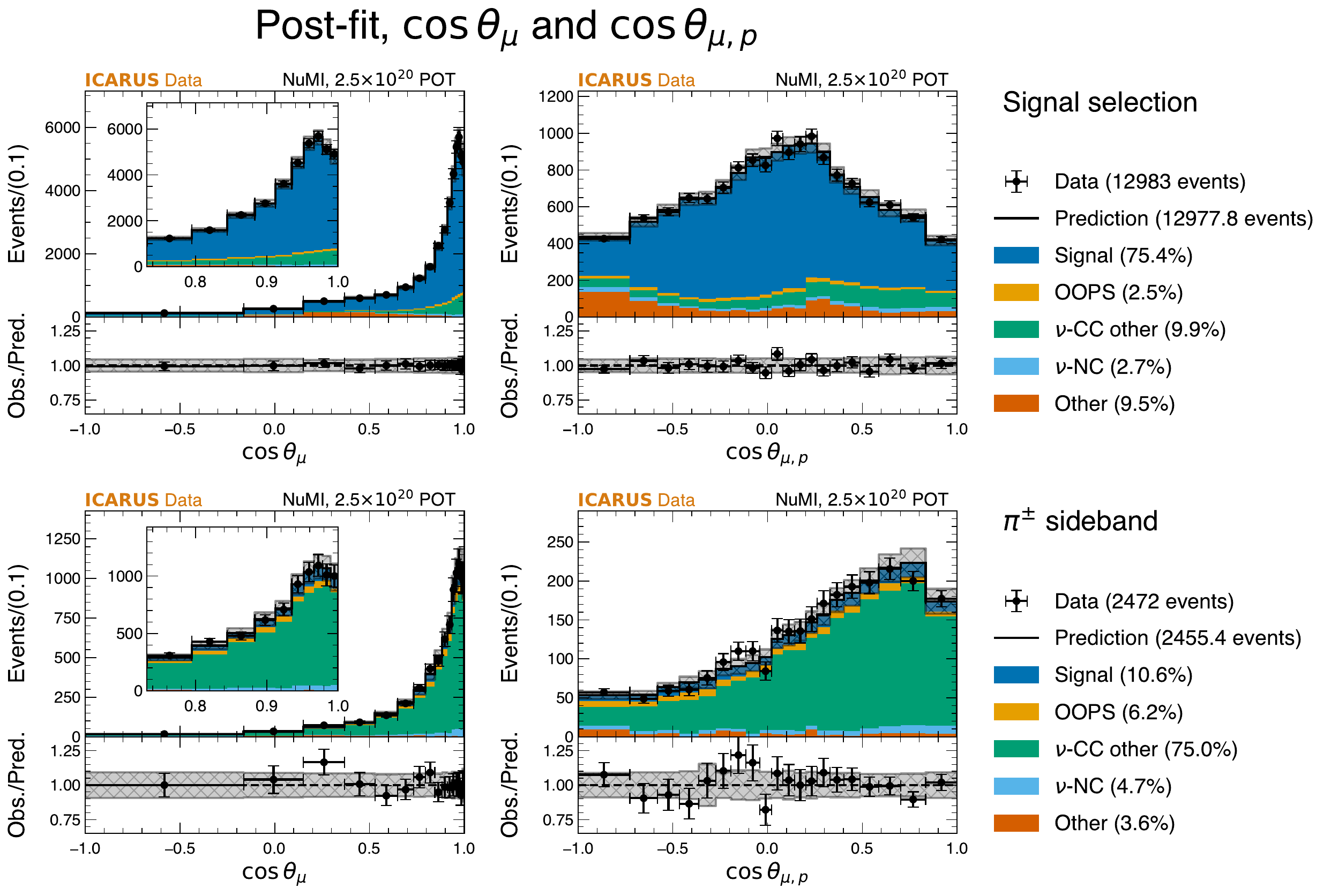}
    \caption{Reconstructed $\cos{\theta_\mu}$ (left) and $\cos{\theta_{\mu,p}}$ (right) distributions after the fit (``post-fit"). Events that pass signal (sideband) selection described in Sec.~\ref{sec:RecoSel:EvSel} (\ref{sec:RecoSel:Sideband}) are shown in the top (bottom) plots. The gray hatched bands represent the statistical and systematic uncertainties on the prediction.}
    \label{fig:PostfitHist_Angular}
\end{figure*}

Once we obtain the best-fit parameters, scaling one set of template parameters by a constant and dividing the other set by the same constant results in the same rate of signal events.
This degeneracy in the template parameters makes the fitter unable to calculate the uncertainties through calculating the Hessian matrix.
To break such degeneracy, we introduced a regularization.
The idea is based on the fact that the template parameters will stay at the value of $1.0$ for the ideal case where we have a perfect modeling.
For each variable, the following penalty term is added to the LLH:
\begin{align}
\label{eq:LLHTemplateSumRegCov1}
\mathrm{LLH}_{\mathrm{Reg.}} &= \lambda_{\mathrm{Reg.}} \left( N - \sum_{i=1}^{N}c_{i} \right)^{2} = \lambda_{\mathrm{Reg.}} \left( \sum_{i=1}^{N}(c_{i}-1) \right)^{2} \\
& = \lambda_{\mathrm{Reg.}} \left[ (c_{1}-1) + (c_{2}-1) + \cdots + (c_{N}-1) \right]^{2} \\
& =\lambda_{\mathrm{Reg.}} (\vec{c} - \vec{c}_{\mathrm{Prior}})^{T}\boldsymbol{C}^{-1}_{\mathrm{Reg.}}(\vec{c} - \vec{c}_{\mathrm{Prior}}),
\end{align}
where $\lambda_{\mathrm{Reg.}}$ controls the strength of the regularization, 
$\vec{c} = \begin{pmatrix} c_{1} & c_{2} & \cdots c_{N} \end{pmatrix}^{T}$ is the template parameters,
$\vec{c}_{\mathrm{Prior}} = \begin{pmatrix} 1.0 & 1.0 & \cdots 1.0 \end{pmatrix}^{T}$ is the prior of the template parameters which are set to 1.0,
and 
\begin{align}
\label{eq:LLHTemplateSumRegCov2}
\boldsymbol{C}^{-1}_{\mathrm{Reg.}} &= \begin{bmatrix}
1 & 1 & \cdots & 1 \\
1 & 1 & \cdots & 1 \\
\vdots & \ddots & \cdots & 1 \\
 1 & 1 & \cdots & 1 \\
\end{bmatrix}
\end{align}
is the inverse of the covariance matrix that defines the regularization.
The strength $\lambda_{\mathrm{Reg.}}$ can be optimized from the ``L-curve"~\cite{doi:10.1137/1034115};
for a wide range of $\lambda_{\mathrm{Reg.}}$, perform fits and plot $\mathrm{LLH}_{\mathrm{Reg.}}/\lambda$ versus $\mathrm{LLH}_{\mathrm{Stat.}}$ to find the point with the largest curvature.
The optimization process is shown in Appendix~\ref{sec:appd_lcurve}. The value of $\lambda_{\mathrm{Reg.}}$ for the angular and TKI variables fits are chosen to 0.065 and 0.077, respectively.
Figure~\ref{fig:PostfitHist_Angular} and \ref{fig:PostfitHist_TKI} show the post-fit distributions on ($\cos{\theta_{\mu}}$, $\cos{\theta_{\mu,p}}$) and ($\delta p_{T}$, $\delta \alpha_{T}$), respectively.

\begin{figure*}
    \centering
    \includegraphics[width=0.90\textwidth]{plots/GUNDAMPlots/SimFit/deltaPT_and_deltaalphaT/post_Histogram_All.pdf}
    \caption{Reconstructed $\delta p_{T}$ (left) and $\delta \alpha_{T}$ (right) distributions after the fit (``post-fit"). Events that pass signal (sideband) selection described in Sec.~\ref{sec:RecoSel:EvSel} (\ref{sec:RecoSel:Sideband}) are shown in the top (bottom) plots. The gray hatched bands represent the statistical and systematic uncertainties on the prediction.}
    \label{fig:PostfitHist_TKI}
\end{figure*}

Mock data sets were generated from the pre-fit covariance matrix to evaluate $p$-values of the data fits.
The procedures and the results are summarized in Appendix.~\ref{sec:appd_pvaluetest}, and all of the $p$-values are above 0.05.

\subsection{Extraction}
\label{sec:xsecextraction:extraction}

The covariance matrix between the fit parameters are used to generate toys of parameters using the Cholesky decomposition of the matrix.
As described in Appendix~\ref{sec:appd_gundam_xsecextraction}, the efficiency correction is effectively applied by propagating the toy parameters to the all true signal events rather than the signal events that are selected.
For each set of toy parameters, a bin-by-bin flux-averaged cross-section is evaluated:
\begin{align}
\left< \frac{d\sigma}{dx} \right>_{\text{Flux}} &= \frac{N_{\text{Signal}}}{\Delta x \cdot \Phi \cdot N_{\text{Target}}}.
\end{align}
$N_{\text{Signal}}$ is the number of signal events with the toy parameters propagated, and $\Delta x$ is the width of the bin.
$\Phi$ is obtained by adding the integrals of the $\nu_{\mu}$ and $\bar{\nu}_{\mu}$ fluxes in Fig.~\ref{fig:ICARUSFlux} and multiplying the result by the POT.
The number of argon nuclei in the fiducial volume, $N_{\text{Target}}$, is calculated from the fiducial volume of liquid argon (Sec.~\ref{sec:RecoSel:EvSel}) using a density of 1.39\,$\text{g}/\text{cm}^{3}$ combined with the atomic mass of argon.
The distributions of $\left< \frac{d\sigma}{dx} \right>_{\text{Flux}}$ of toys are used to evaluate uncertainty and the correlation between the cross-section bins.
The multiplication of two log-normal distributions follows a log-normal distribution, so the covariance matrix is calculated in log-scale. 
The uncertainties are taken as the one standard deviation in log-scale which results in asymmetric errors in the extracted cross section in linear-scale.

\begin{figure*}
    \centering
    \includegraphics[width=0.90\textwidth]{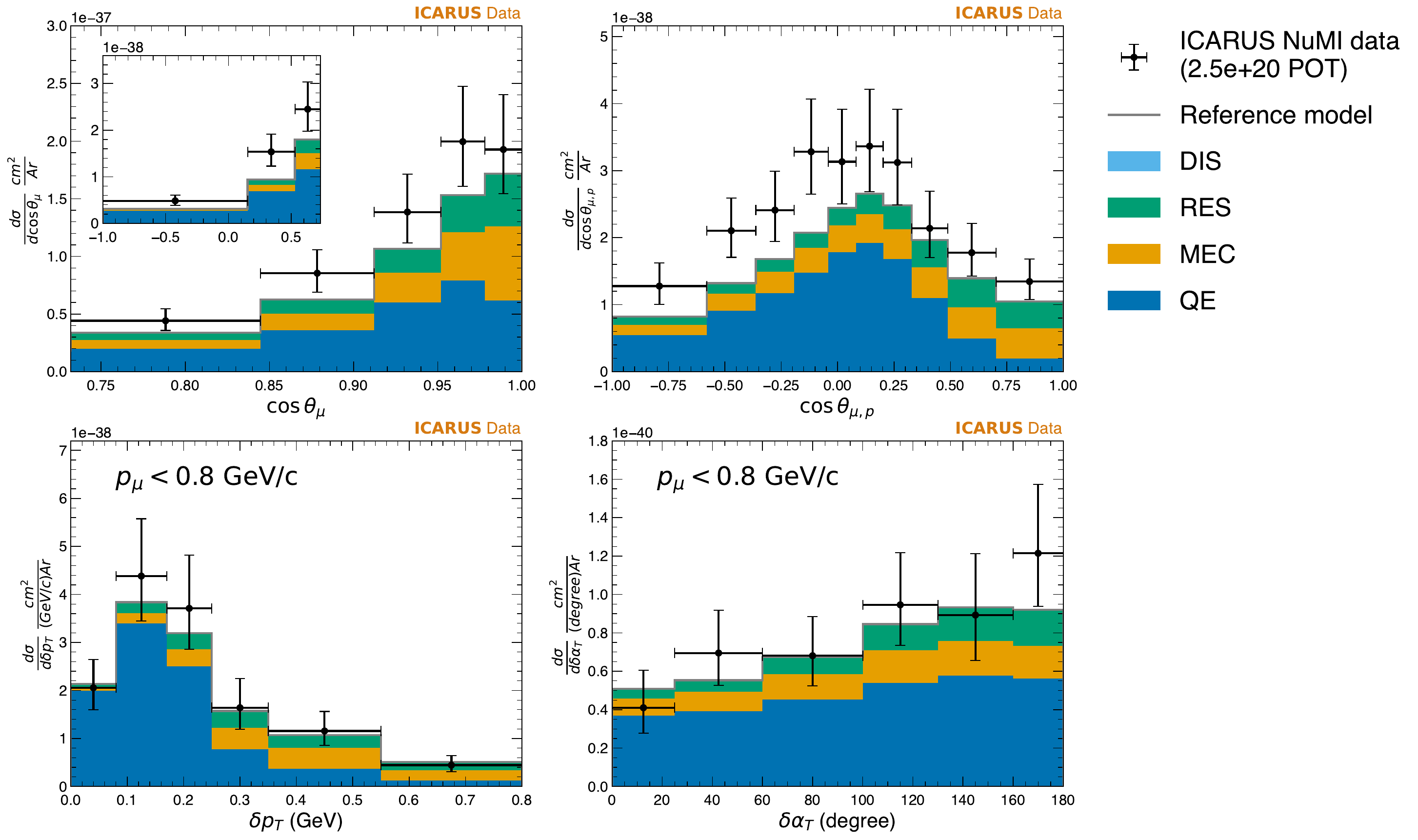}
    \caption{The extracted cross sections (black marker) compared to reference GENIE model (\texttt{AR23\_20i}) prediction broken down into interaction mode (QE, MEC, RES and DIS). As described in Sec.~\ref{sec:RecoSel:SigDef} and \ref{sec:RecoSel:EvSel}, the signal events used to extract cross sections as a function of TKI variables (bottom) are restricted to have muon momentum below $0.8$\,GeV/c.}
    \label{fig:xsec_AnaBin_GENIEBreakDown}
\end{figure*}

As described in Sec.~\ref{sec:eventsim:nugenerator}, \GENIE \texttt{AR23\_20i} CMC is used as the reference model.
Figure~\ref{fig:xsec_AnaBin_GENIEBreakDown} shows the extracted cross sections for the angular and TKI variables, respectively, with the breakdown of interaction modes predicted by the reference model.
The fractional contribution of each interaction mode with and without the muon momentum restriction is summarized in Table~\ref{tab:GENIEXsecModeFrac}.
The signal is dominated by QE followed by MEC and RES. DIS contributes less than 0.1\%.

\begin{table}[htbp]
    \caption{Fractional contributions of each interaction mode to the signal events predicted from \GENIE reference model (\texttt{AR23\_20i} CMC.)}
    \centering
    \label{tab:GENIEXsecModeFrac}
    \begin{tabular}{c|cc}
\multirow{2}{*}{Mode} & \multirow{2}{*}{All signal events} & Signal events with\\
 & & $p_\mu<0.8~\mathrm{GeV/c}$ \\
\hline
QE & 59.9\% & 65.6\% \\ 
MEC & 22.4\% & 19.1\% \\
RES & 17.6\% & 15.2\% \\
DIS & \multicolumn{2}{c}{$<0.1$\%} \\
    \end{tabular}
\end{table}

\subsection{Fractional uncertainties on the cross section}
\label{sec:fracuncxsec}

\begin{figure*}
    \centering
    \includegraphics[width=0.90\textwidth]{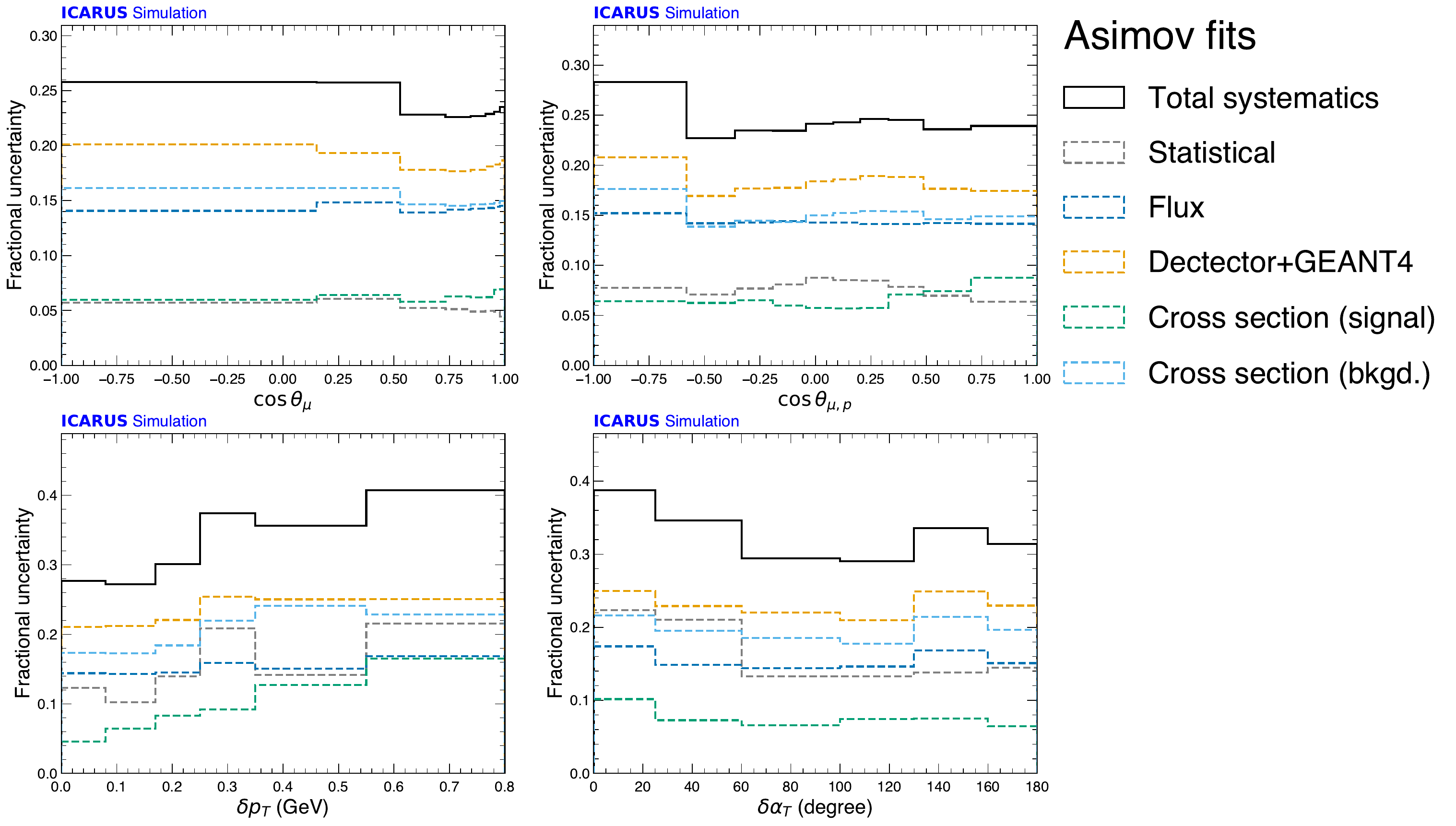}
    \caption{An illustration of fractional uncertainties of each group of systematic uncertainty from the ``N-minus-X method." The cross-section uncertainties that are related to QE, MEC, and FSI modeling are grouped into ``Cross section (signal)" category, and the rest are included in ``Cross section (bkgd.)."}
    \label{fig:xsec_AnaBin_FracUnctFromNminusX}
\end{figure*}

The \GUNDAM workflow involves a fit prior to the extraction step, hence the fractional uncertainties that are typically reported together with the cross-section results are not trivially accessible in this analysis.
A similar quantity can be evaluated by performing the fit and extraction on an Asimov dataset but with one set of systematic uncertainties excluded (``N-minus-X method", where N stands for all uncertainties and X stands for the ones that are excluded).
The set of all systematic uncertainties were grouped into ``categories'' based on their origin (Flux, Detector+\GEANTfour, signal cross-section uncertainties, and background cross-section uncertainties), and the N-minus-X procedure is performed for each category. 
The uncertainties are compared to the results with full systematics uncertainties included (N), and the square-root of the squared-difference is considered as the fractional uncertainty of the set of systematic uncertainties excluded (X). The statistical uncertainties can be obtained in a similar manner including signal template parameters without any systematic uncertainties. The result of this method is presented in Fig.~\ref{fig:xsec_AnaBin_FracUnctFromNminusX}.

%% file: Sections/Results.tex
\section{Results and comparisons with Generators}
\label{sec:ResultAndComparison}

Alterative \GENIE samples are generated with the same configuration as \texttt{AR23\_20i} CMC but with the QE axial form factor ($F_{A}^{CCQE}$) choices reported in Ref.~\cite{meyer2025nucleonaxialformfactor}.
The first set uses $z$-expansion parameters fitted to an average of various LQCD calculations~\cite{meyer2026nucleonaxialformfactor}  (``LQCD $z$-exp $F_{A}^{CCQE}$").
The LQCD calculations typically predict higher values of $F_{A}^{CCQE}$ for higher values of $Q^{2}=-q^{2}$, where $q$ is the four-momentum transfer of the scattering,
than the $z$-expansion fit on the bubble chamber data~\cite{PhysRevD.93.113015} which is used in \texttt{AR23\_20i} CMC.
The second set uses $z$-expansion parameters fitted to the results mearusred by MINERvA collaboration using the antineutrino–proton scattering data~\cite{MINERvA:2023avz} (``MINERvA $z$-exp $F_{A}^{CCQE}$").

\GENIE provides various models to simulate FSI. 
The two popular choices are \texttt{hA2018} and \texttt{hN2018}~\cite{Dytman:2011zz,PhysRevD.104.053006}.
The former, where the FSI is simulated at the hadron-nucleus level, provides systematic variations through a reweighting scheme and is the default model chosen in \texttt{AR23\_20i}.
The latter (``hN2018 FSI") allows multiple cascades of the initial hadron inside the nucleus.

Several alternative event generators are also compared to the extracted cross section.

\NUWRO~\cite{GOLAN2012499} version 25.11 is used to provide an alternative prediction with the argon spectral function measured by the Jefferson Lab Hall A Collaboration~\cite{PhysRevD.105.112002}.
The spectral function captures the detailed shell structure of the argon nucleus.
The Llwellyn-Smith model~\cite{LLEWELLYNSMITH1972261} is used for the QE interaction, with the $F_{A}$ using the $z$-expansion fit result from the MINERvA measurement~\cite{MINERvA:2023avz}.

\GIBBU~\cite{Buss:2011mx} has unique treatment of particle transport inside the medium through the  Boltzmann-Uehling-Uhlenbeck (BUU) equation~\cite{BBlattel_1993}. 
We used the \texttt{r2025\_02} release of \GIBBU with the isospin of the argon ($T$) set to $1$.

\NEUT version 6.1.3~\cite{NEUT} is used to provide a comparison to an alternative CCQE model based on the relativistic distorted wave impulse approximation (RDWIA)~\cite{Gonzalez_Jimenez_2020,PhysRevC.105.025502,edrmf}.
The initial nuclear bound state is described using momentum distributions derived from relativistic mean field (RMF) theory~\cite{Serot:1984ey}.
The implementation in \NEUT incorporates a missing energy profile parameterized by Gaussian function~\cite{edrmf}. This allows for the user to modify the shell structure of the target nucleus.
The final state is modeled using an energy-dependent relativistic mean field (EDRMF) potential, which modifies the energy-independent RMF potential with a blending function tuned to electron-scattering data.
This approach ensures consistency between the initial and final states, and naturally incorporates the Pauli exclusion principle within a quantum mechanical framework.
FSI is modeled by combining these distorted waves with the standard \NEUT intranuclear cascade, which handles inelastic processes.

\begin{figure*}
    \centering
    \includegraphics[width=0.90\textwidth]{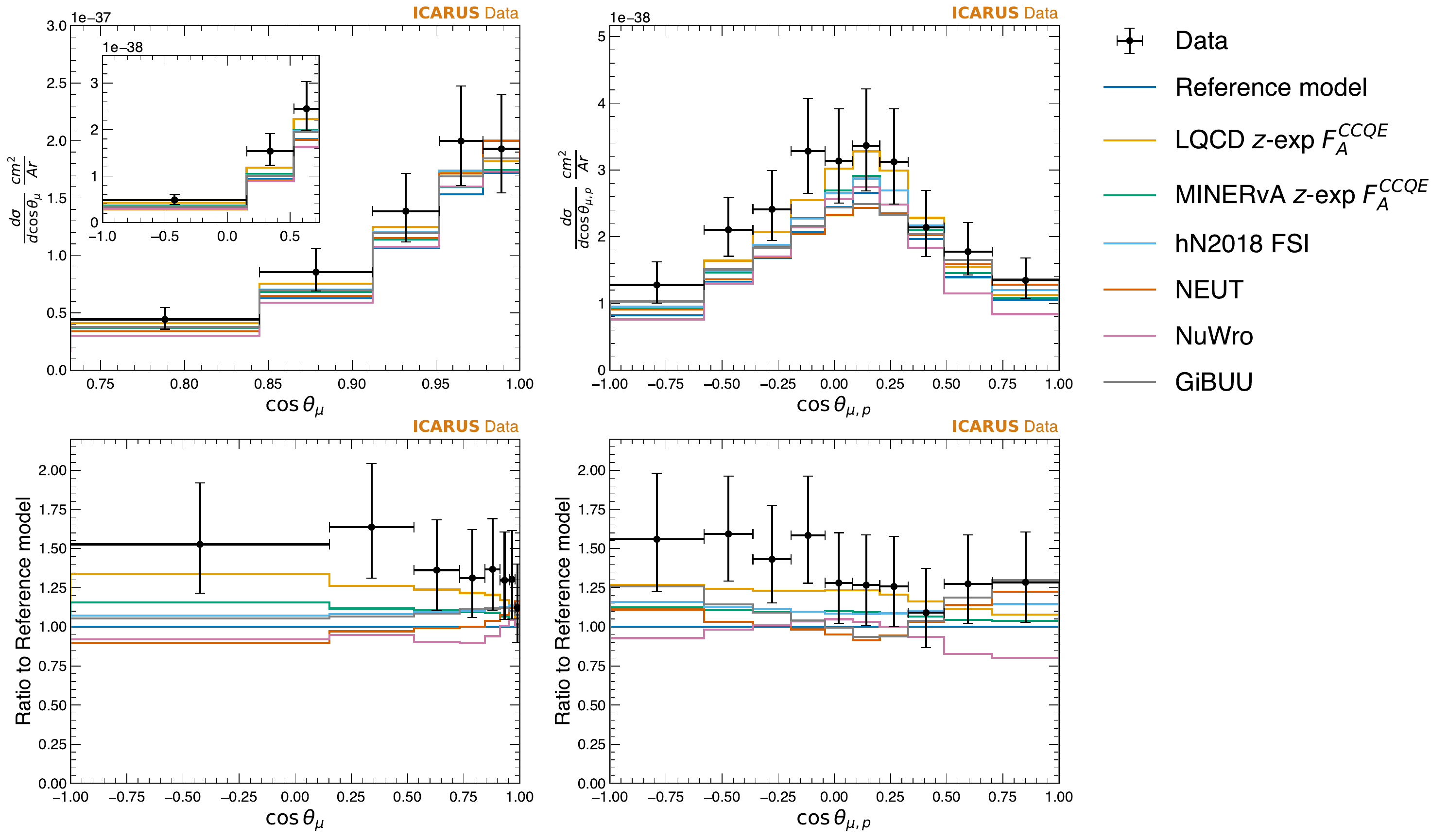}
    \caption{The extracted cross section on $\cos{\theta_\mu}$ and $\cos{\theta_{\mu,p}}$ (black marker) compared to various model predictions (top) and their ratio to the reference \GENIE model (\texttt{AR23\_20i\_00\_000}) (bottom).}
    \label{fig:xsec_AnaBin_Angular_With_RatioToGENIE}
\end{figure*}

\begin{figure*}
    \centering
    \includegraphics[width=0.90\textwidth]{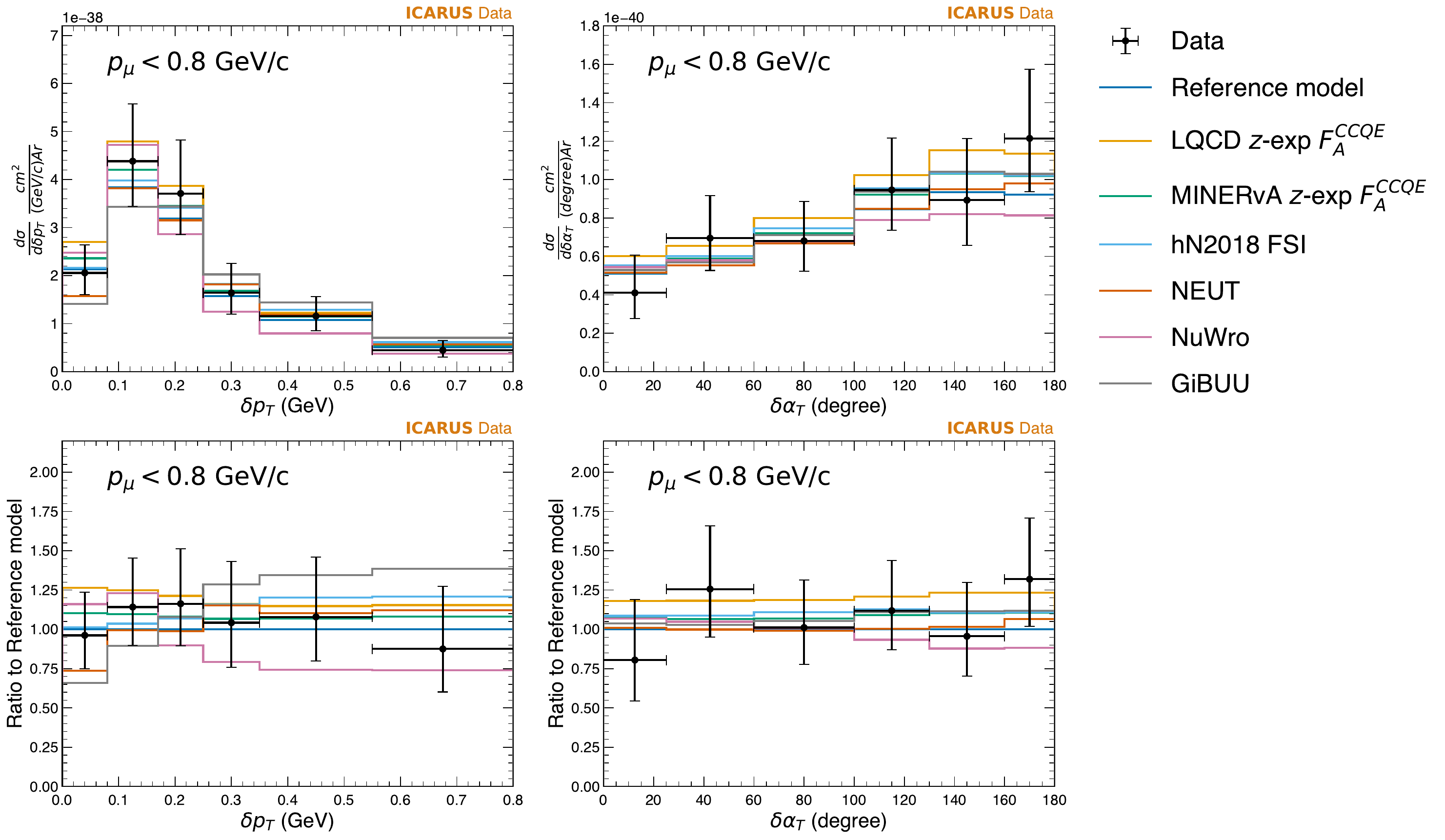}
    \caption{The extracted cross section on $\delta p_{T}$ and $\delta \alpha_{T}$ (black marker) compared various model predictions (top) and their ratio to the reference \GENIE model (\texttt{AR23\_20i\_00\_000}) (bottom).}
    \label{fig:xsec_AnaBin_TKI_With_RatioToGENIE}
\end{figure*}

Figure~\ref{fig:xsec_AnaBin_Angular_With_RatioToGENIE} and \ref{fig:xsec_AnaBin_TKI_With_RatioToGENIE} show the extracted cross section compared to various generator predictions described above.
The covariance matrix from the extraction (Sec.~\ref{sec:xsecextraction:extraction}) is used to evaluate the $\chi^{2}$ and $p$-value for each of the prediction (Table~\ref{tab:xsec_chi2_summary}).

\begin{table}[htbp]
    \caption{The $\chi^{2}$ divided by the degrees of freedom ($\chi^{2}/ndof$) and $p$-value calculated for each model prediction against the extracted cross section.}
    \centering
    \label{tab:xsec_chi2_summary}
    \begin{tabular}{l|cc|cc}
\multirow{2}{*}{Prediction} & \multicolumn{2}{c|}{Angular variables} & \multicolumn{2}{c}{TKI variables} \\
 & $\chi^{2}/ndof$ & $p$-value & $\chi^{2}/ndof$ & $p$-value \\
\hline
Reference model & 13.44/18 & 0.765 & 4.44/12 & 0.974 \\
LQCD $z$-exp $F_{A}^{CCQE}$ & 7.27/18 & 0.988 & 4.58/12 & 0.971 \\
MINERvA $z$-exp $F_{A}^{CCQE}$ & 9.40/18 & 0.950 & 4.15/12 & 0.981 \\
hN2018 FSI & 12.79/18 & 0.804 & 4.89/12 & 0.962 \\
NEUT & 23.02/18 & 0.190 & 5.57/12 & 0.936 \\
NuWro & 17.34/18 & 0.500 & 13.73/12 & 0.318 \\
GiBUU & 15.88/18 & 0.601 & 13.38/12 & 0.342 \\   
    \end{tabular}
\end{table}

The impact of the higher $F_{A}$ from the LQCD calculation and MINERvA measurement can be seen in Fig.~\ref{fig:xsec_AnaBin_Angular_With_RatioToGENIE},
which is expected from the corelation between $Q^{2}$ and scattered angles of the out-going particles.
The nuclear ground state is sensitive to the TKI variables shown in Fig.~\ref{fig:xsec_AnaBin_TKI_With_RatioToGENIE}.
\NEUT shows the lowest $p$-value in the angular variables at 0.190, but this is still compatible within two standard deviations. The $p$-value of \NEUT is higher in the TKI variables (0.936).
\texttt{hN2018} predicts higher cross section at the higher tail of the $\delta p_{T}$ distribution, where the impact of FSI is expected to be most prominent.
The unique shapes predicted by \NUWRO and \GIBBU in the $\delta p_{T}$ distribution in both the Fermi peak (150--200\,MeV) and the higher tail results in the two least compatible predictions in the TKI variables, but still within one standard deviation.
The current budget of both statistical and systematic uncertainties does not provide enough discrimination power between various models that we compared to our extracted cross section (i.e., none of the models are rejected by $p$-value criteria of 5\%).
As illustrated in Fig.~\ref{fig:xsec_AnaBin_FracUnctFromNminusX}, the analysis is limited by both statistical and systematic uncertainties.
The former can possibly be overcome by adding in additional data taken with NuMI with the opposite beam polarity (though this has a somewhat different flux at ICARUS) and/or by collecting more NuMI data at the ICARUS detector in the same beam polarity.
The largest source of systematic uncertainty is coming from the detector, and there are on-going efforts on improving our knowledge of the detector and the reality of the simulation, which will be implemented in the upcoming ICARUS physics analyses.
Some of the detector effects are specific to the ICARUS configuration (such as long horizontal wires and warm electronics readout), and other effects and efforts to characterize them are part of the calibration procedure that are beneficial to the broader LAr TPC program.

%% file: Sections/Conclusion.tex
\section{Conclusion}
\label{sec:concl}

In this paper, we reported the first neutrino-argon cross-section measurement using off-axis NuMI beam data at ICARUS, using $2.5 \times 10^{20}$ protons-on-target (POT).
The results will provide valuable input to improve the modeling of neutrino interactions for future neutrino oscillation experiments such as DUNE, which shares similar energy spectrum and target.
The flux-averaged cross sections are reported for charged-current quasielastic-like events, where the final state consists of a muon, protons, and neutrons without any other mesons or baryons.
Four one-dimensional cross sections are extracted for two angular and two transverse kinematic imbalance (TKI) variables: muon angle with respect to the neutrino direction ($\cos{\theta_{\mu}}$), opening angle between the muon and the leading proton ($\cos{\theta_{\mu,p}}$), $\delta p_{T}$, and $\delta \alpha_{T}$.
For the TKI variables where the contained tracks are used as muon candidates, a phase-space restriction of muon momentum below 0.8 GeV/c is required.
The two angular (TKI) variables are fit simultaneously, hence the correlation between the variables are kept even though the final results are one-dimensional cross sections.
The extracted cross sections are compared to various generator predictions, including the reference model from \texttt{GENIE} with \texttt{AR23\_20i\_00\_000} (``\texttt{AR23\_20i}") comprehensive model configurations (CMC).
The reported cross sections are consistent with the various interesting predictions used for the comparison with the current budget of uncertainties.
Further analysis with more data and improved detector systematics will be crucial in distinguishing among models.

%% file: Sections/Appendix_GUNDAM_Exraction.tex
\section{Cross-section extraction using GUNDAM package}
\label{sec:appd_gundam_xsecextraction}

\begin{figure}[htbp]
    \centering
    \includegraphics[width=0.45\textwidth]{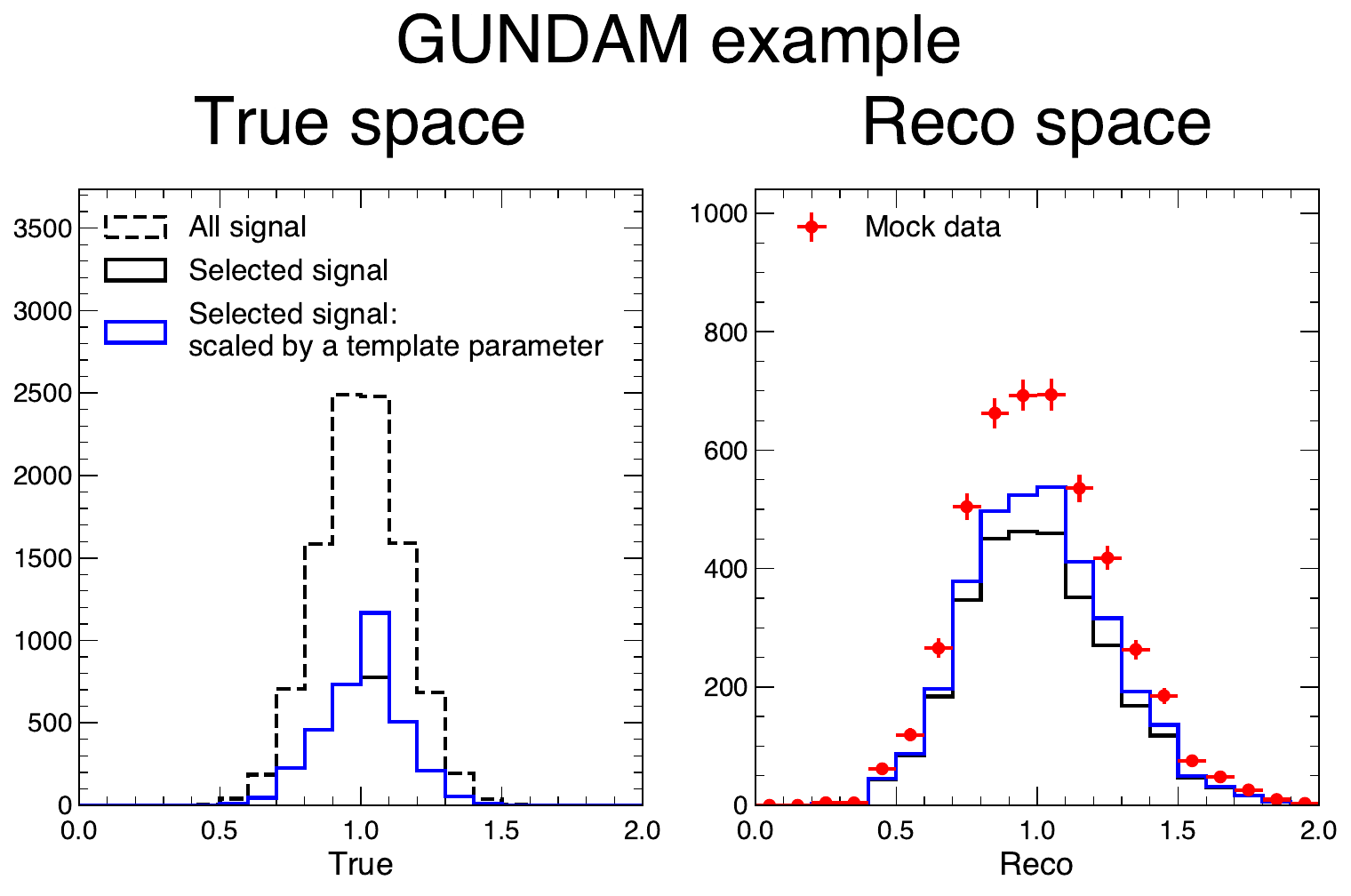}
    \caption{A simple example experiment is introduced to describe the cross-section extraction through the \GUNDAM~\cite{GUNDAMRepo} package. The true distribution on the left is generated from a Gaussian distribution, with the mean and standard deviation of 1.0 and 0.15. A flat selection efficiency of 30\% is assumed and used to decide whether an event is selected or not by comparing a random number thrown from a uniform distribution to 0.30. A template parameter assigned for the true bin of $[1.0, 1.1]$ is set to 1.5, which increases the same bin by 50\%. We assumed a 20\% smearing in the reconstruction of this true variable. The impact of the template parameter is smeared into multiple bins in the reconstructed (Reco) space on the right. ``Mock data" on the right is configured by reweighting each event by $1 + 0.5 \times \mathrm{True}$.}
    \label{fig:GUNDAMExample1}
\end{figure}

\begin{figure}[htbp]
    \centering
    \includegraphics[width=0.45\textwidth]{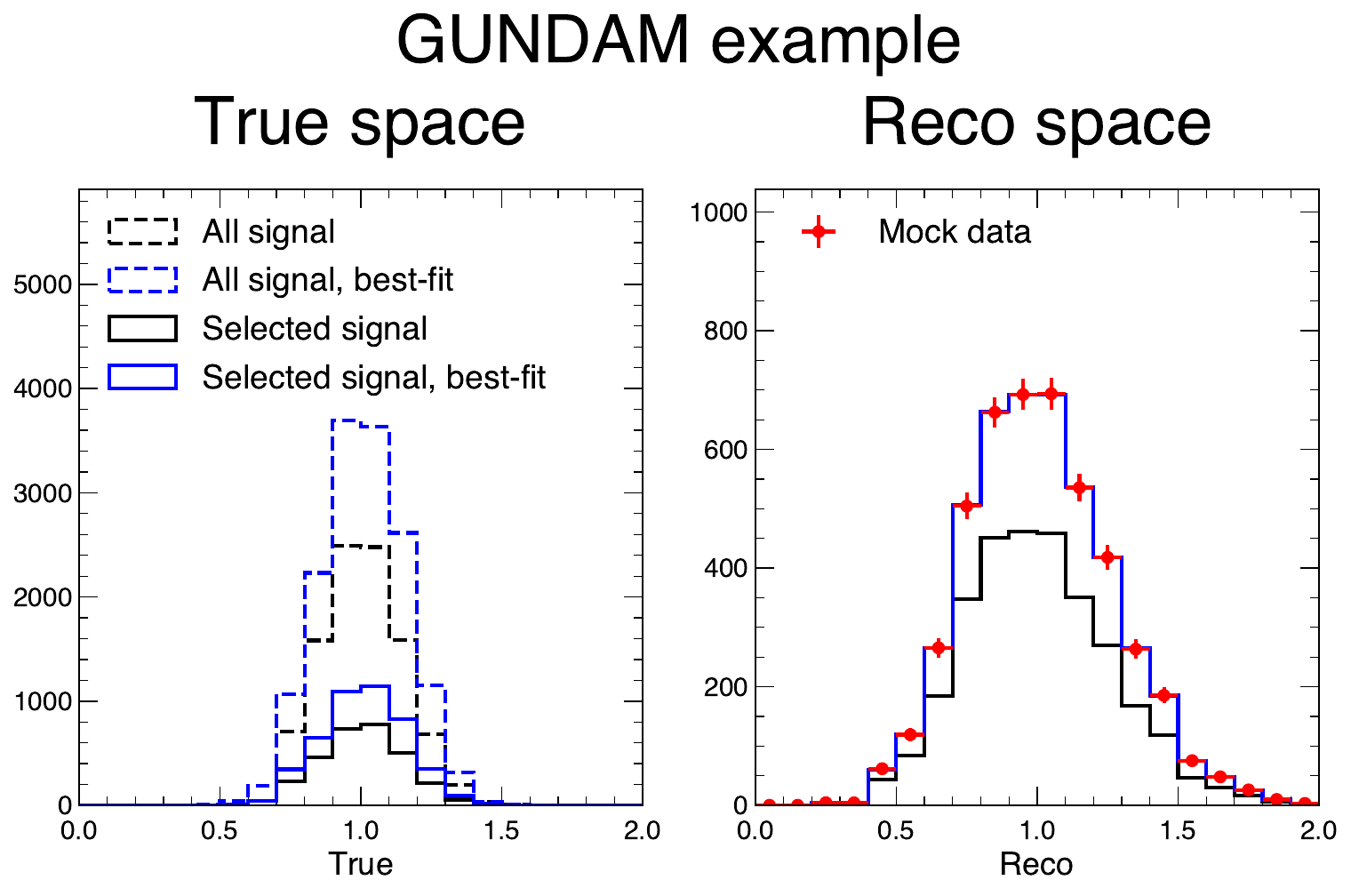}
    \includegraphics[width=0.45\textwidth]{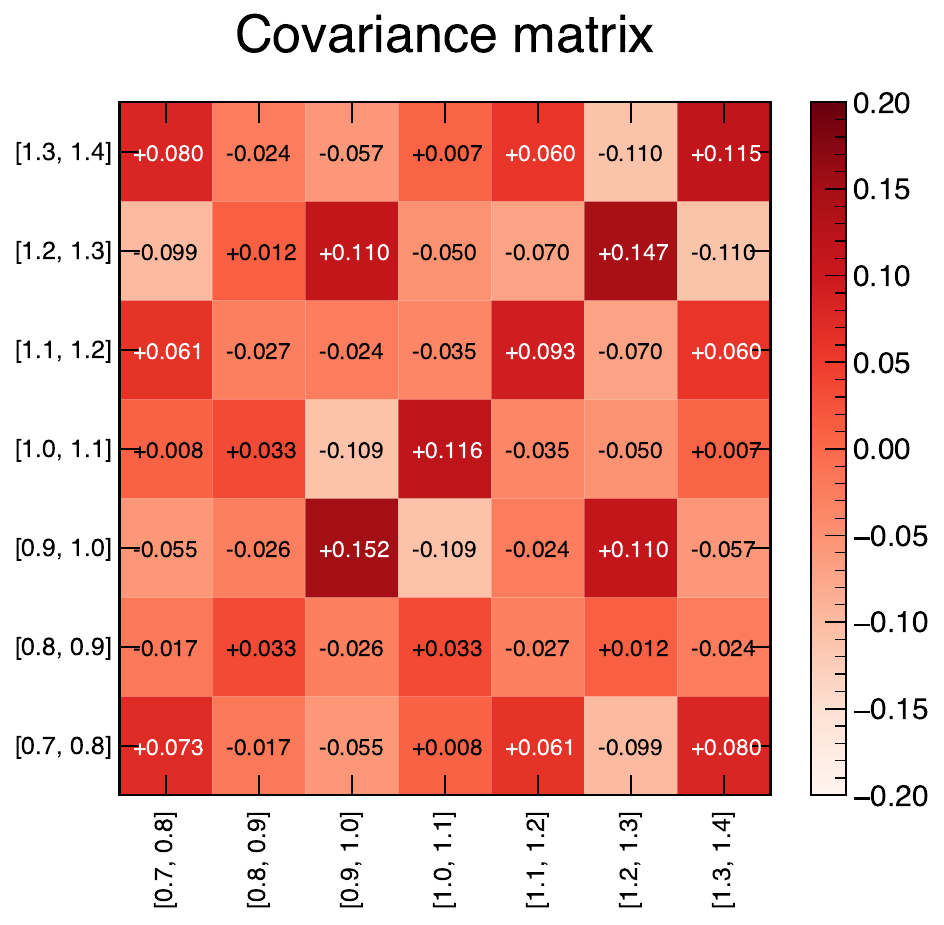}
    \caption{A fit is performed to find the set of template parameters that minimizes the log-likelihood, $-2\log{\mathcal{L}} = 2 \sum_{\text{Reco}} \left( M - D + D \times \log{D/M} \right)$, where $M$ and $D$ represent the predicted and observed (``Mock data") number of events in each bin of reconstructed space in Fig.~\ref{fig:GUNDAMExample1}. The covariance matrix of the template parameters is also shown.}
    \label{fig:GUNDAMExample2}
\end{figure}

In this section, we illustrate how the unfolding, efficiency correction, and cross-section extraction is performed by \GUNDAM with an example. On the left plot of Fig.~\ref{fig:GUNDAMExample1}, a true distribution of all signal is drawn in the black dashed line.
Assuming a flat 30\% selection efficiency, the same distribution but only for selected signal events is drawn in the black solid line.
We also assume a 20\% Gaussian smearing in the reconstruction.
The goal of unfolding is to obtain a ``true" distribution from the reconstructed quantities.
The template parameters are defined as the scale factors applied to signal events for each true variable bin.
The blue histogram is same as the selected signal distribution  (black solid line) but with a template parameter of 1.5 assigned to the $\left[1.0, 1.1\right]$ bin.
The impact of the template parameter is distributed in multiple bins in the reconstructed (``Reco") space on the right plot of Fig.~\ref{fig:GUNDAMExample1}.
Mock data are made by reweighting each event by $1 + (0.5 \times \mathrm{True value})$; a linear scaling as a function of the True value that has no modification at True value of 0, doubles the simulated counts at True value of 2, etc.
A fit is performed by varying parameters that affect the reco distribution, and finds the combination that minimizes the log-likelihood (LLH), $-2\log{\mathcal{L}}$, where $\mathcal{L}$ is the likelihood and $-2$ is multiplied so that the LLH then asymptotically approaches $\chi^{2}$ distribution~\cite{10.1214/aoms/1177732360}.
In the simple example described above, we have template parameters for each true bin as free parameters without any systematic uncertainties, and a Poisson likelihood is used: $\mathrm{LLH}_{\mathrm{Poisson}} = 2 \sum_{\text{Reco}} \left( M - D + D \times \log{D/M} \right)$ where $M$ and $D$ represent predicted and observed number of events for each bin in the reconstructed space, respectively.
The fit result of the example is shown in Fig.~\ref{fig:GUNDAMExample2}.
The template parameters are extracted along with the correlation between them.

\begin{figure}[htbp]
    \centering
    \includegraphics[width=0.40\textwidth]{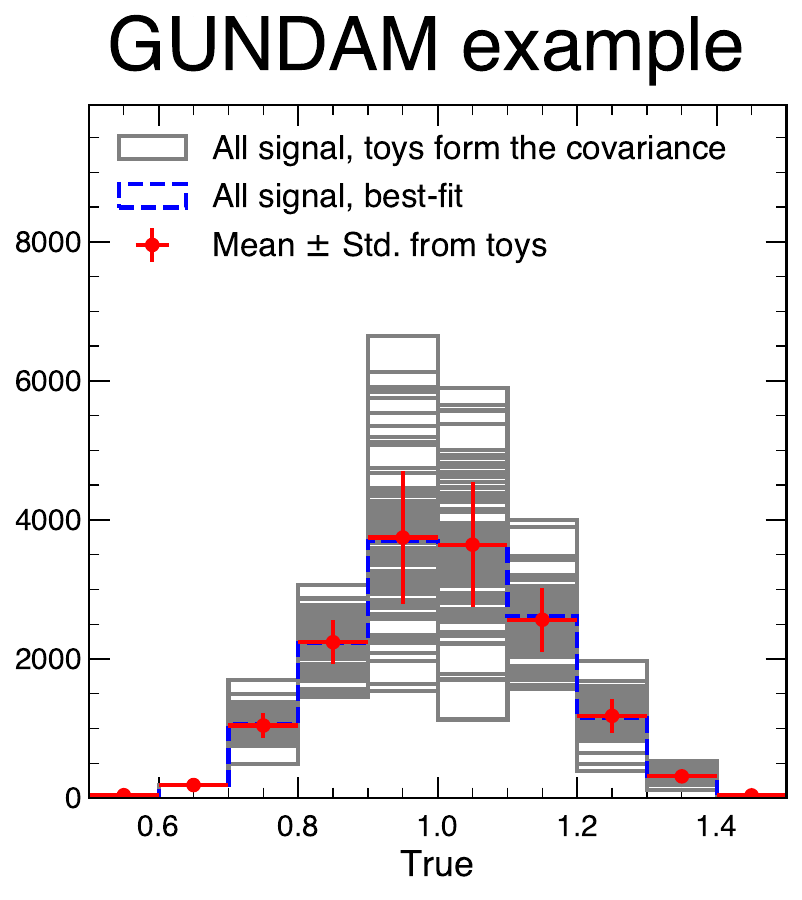}
    \caption{The post-fit covariance matrix of the fit parameters from Fig.~\ref{fig:GUNDAMExample2} is used to generate toys of parameters. Each toy is then applied to the true distribution of all signal events (gray). The mean and standard deviation of the toys are calculated for each true bin (red).}
    \label{fig:GUNDAMExample3}
\end{figure}

The blue solid line in the top left plot in Fig.~\ref{fig:GUNDAMExample2} is the unfolded distribution that best describes the mock data in the reco distribution on the right plot.
By construction, the fitting is done only using the events that are selected.
The ``efficiency correction" can be done by evaluating the efficiency and divide the unfolded distribution with it.
However each division requires a proper propagation of fit parameters to both numerator and denominator.
An alternative way of doing the efficiency correction is to use the ``all signal" events rather than the selected signal events which effectively does the efficiency correction.
This is the choice followed within \GUNDAM.
The blue dashed line in Fig.~\ref{fig:GUNDAMExample2} is the distribution of all signal events with the best-fit template parameters applied, and this corresponds to the ``cross section" but without additional normalization such as the division by the number of targets and flux integral.
The uncertainty of the efficiency-corrected unfolded distribution can be obtained by generating ``toys" of fit parameters from the covariance matrix.
Figure~\ref{fig:GUNDAMExample3} shows the best-fit and toy distributions of the true distribution from all signal events.
The bin-by-bin mean and standard deviation (std) of the toys can be used as the central value and error of the extracted ``cross section" of this example.

%% file: Sections/Appendix_pvalue.tex
\section{$p$-value test}
\label{sec:appd_pvaluetest}

\begin{figure}[htbp]
    \centering
    \includegraphics[width=0.45\textwidth]{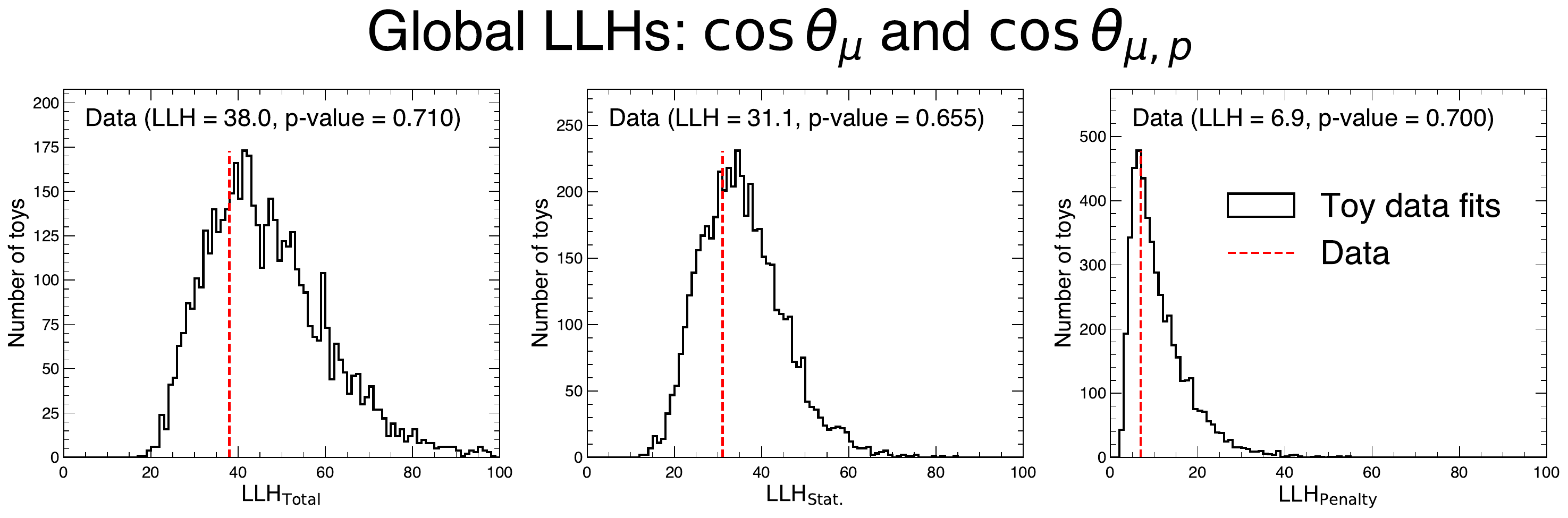}
    \includegraphics[width=0.45\textwidth]{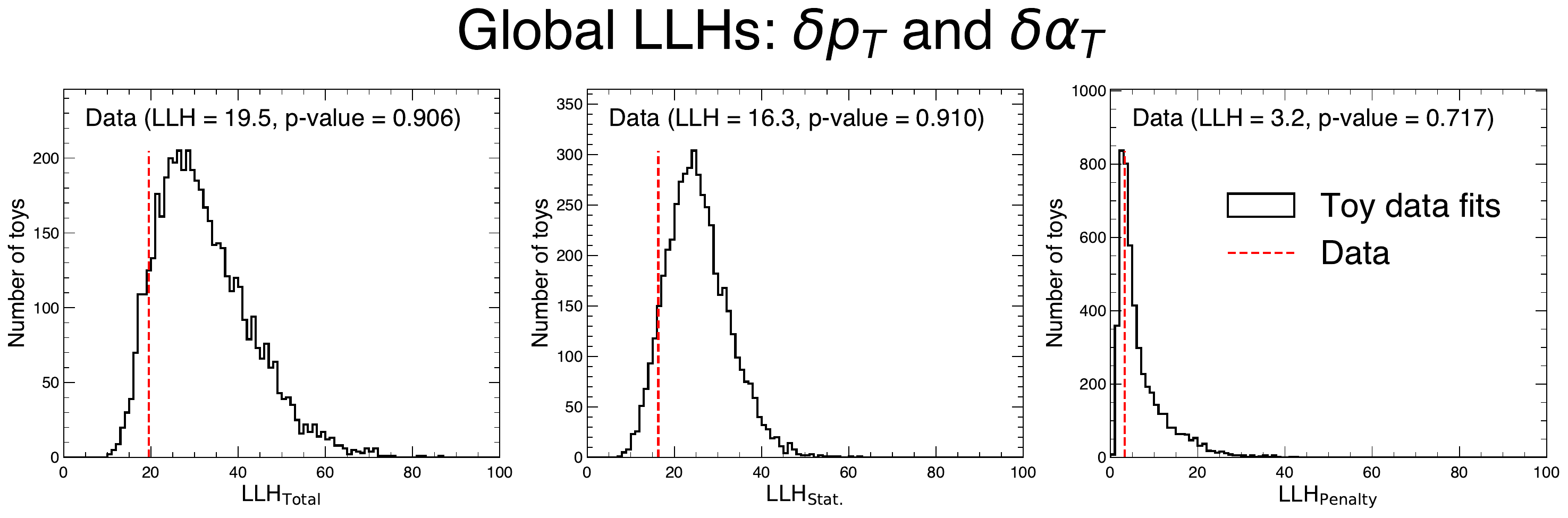}
    \caption{The global LLH distributions from the angular (top) and TKI (bottom) variables.}
    \label{fig:pvalue_Global}
\end{figure}

To validate the fit configuration and the results, sets of mock data were generated from the pre-fit covariance matrix and fit (``toy fits").
The $p$-values are calculated from  various LLHs (described in Sec.~\ref{sec:xsecextraction:fitting}) of the toy fits.
In Fig.~\ref{fig:pvalue_Global}, three ``global" LLH distributions are shown; the $\mathrm{LLH}_{\mathrm{Stat.}}$ including both signal and sideband selected events, $\mathrm{LLH}_{\mathrm{Penalty}}$ from all systematic uncertainties, and the sum of the two ($\mathrm{LLH}_{\mathrm{Total}}$), on angular and TKI variables, respectively.
The observed values from the data fit are indicated by the red dashed vertical lines, and the $p$-value calculated as the fraction of toys equal to or above the observed value.

\begin{figure}[htbp]
    \centering
    \includegraphics[width=0.45\textwidth]{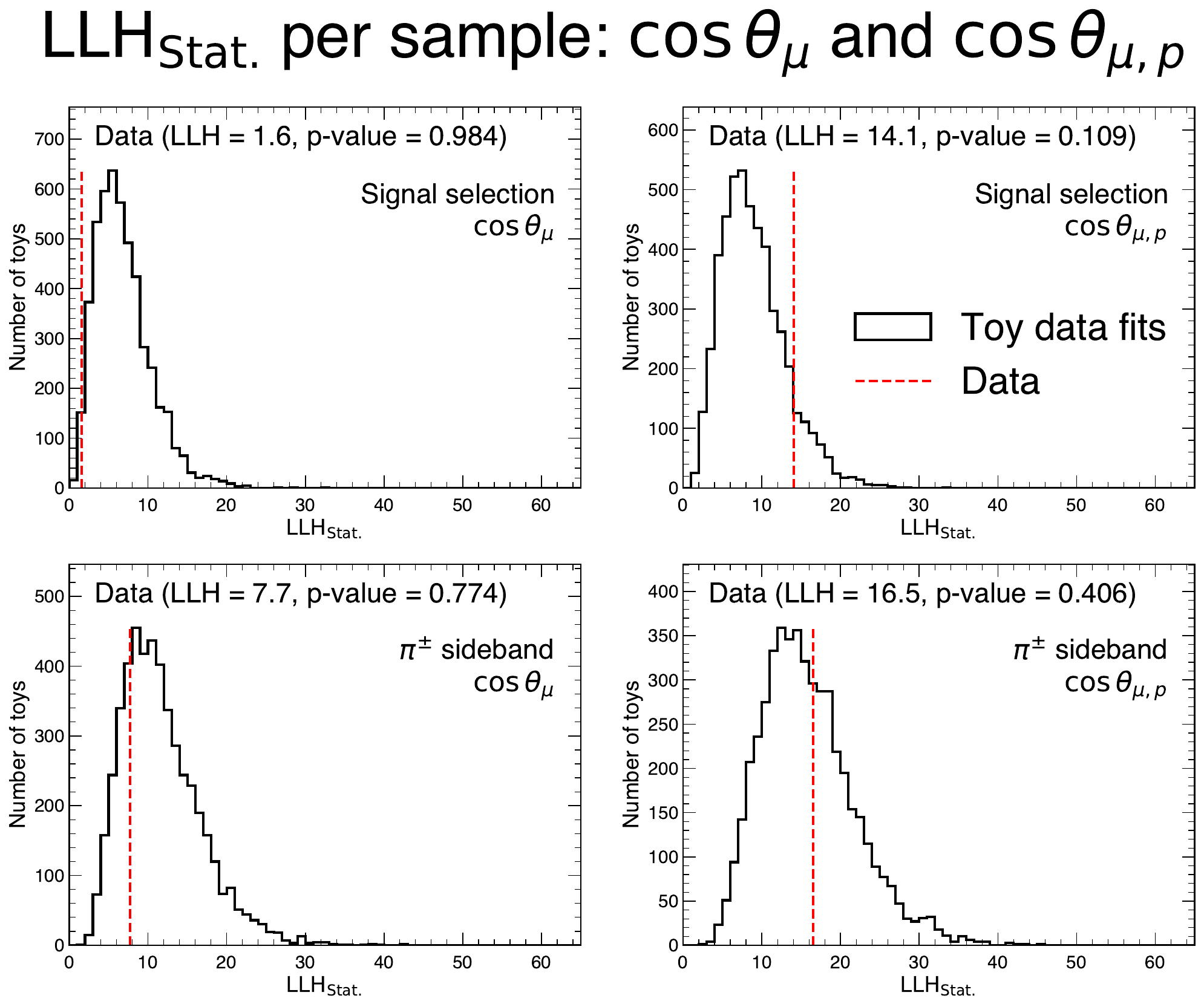}
    \caption{The statistical LLH distributions from the $\cos{\theta_\mu}$ and $\cos{\theta_{\mu,p}}$ fit, calculated separately for each sample.}
    \label{fig:pvalue_PerSample_Angular}
\end{figure}

\begin{figure}[htbp]
    \centering
    \includegraphics[width=0.45\textwidth]{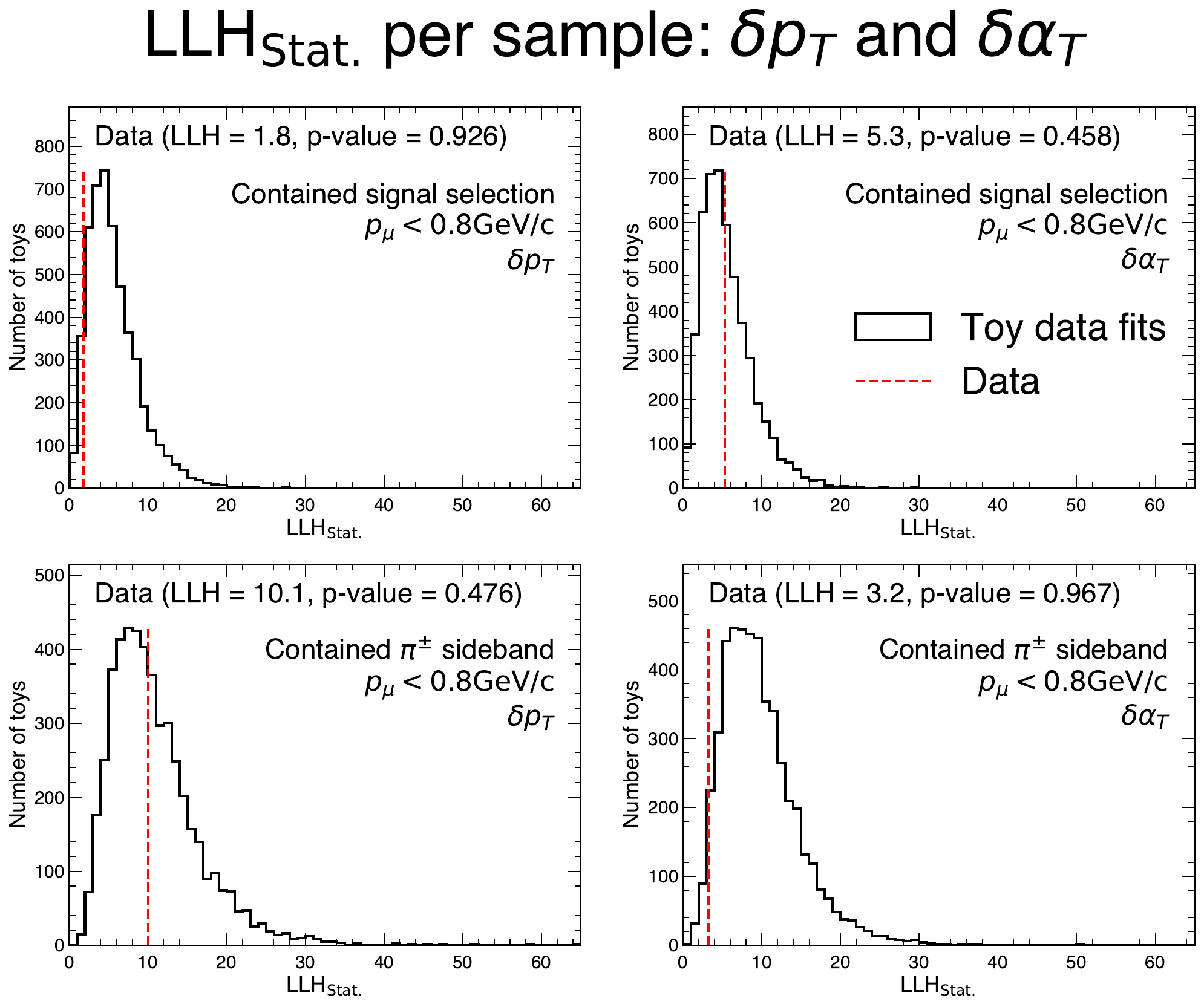}
    \caption{The statistical LLH distributions from the $\delta p_{T}$ and $\delta \alpha_{T}$ fit, calculated separately for each sample.}
    \label{fig:pvalue_PerSample_TKI}
\end{figure}

Figures~\ref{fig:pvalue_PerSample_Angular} and \ref{fig:pvalue_PerSample_TKI} show the $\mathrm{LLH}_{\mathrm{Stat.}}$ distributions calculated independently for each sample. These include the signal selection and $\pi^{\pm}$ sideband, evaluated across both angular and TKI variables.

\begin{figure}[htbp]
    \centering
    \includegraphics[width=0.45\textwidth]{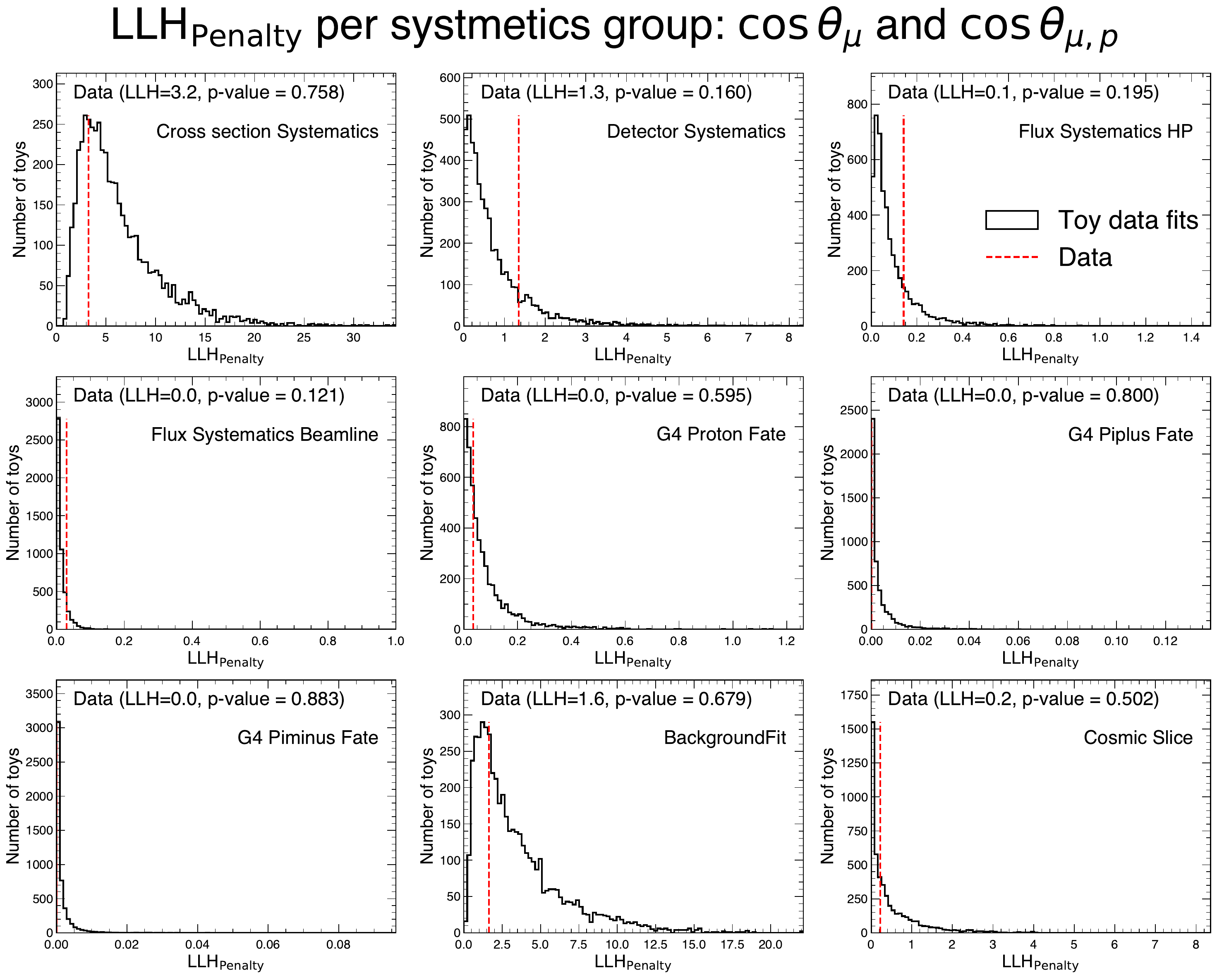}
    \caption{The penalty term of the LLH for each group of systematic uncertainties from the $\cos{\theta_\mu}$ and $\cos{\theta_{\mu,p}}$ fit.}
    \label{fig:pvalue_PerSyst_Angular}
\end{figure}

The final set of LLH is $\mathrm{LLH}_{\mathrm{Penalty}}$ evaluated for each group of systematics and shown in  Fig.~\ref{fig:pvalue_PerSyst_Angular}.

\begin{figure}[htbp]
    \centering
    \includegraphics[width=0.45\textwidth]{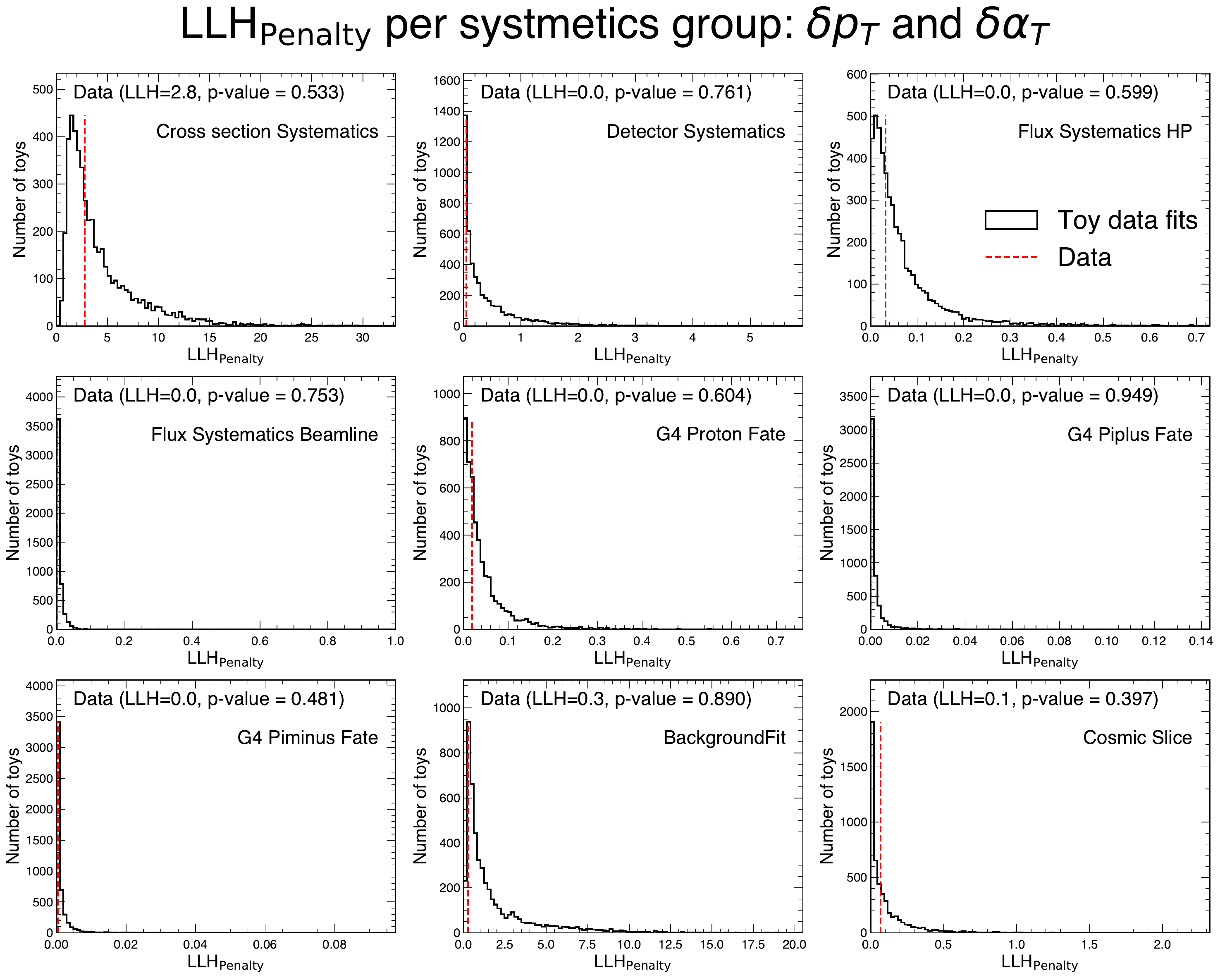}
    \caption{The penalty term of the LLH for each group of systematic uncertainties from the $\delta p_T$ and $\delta \alpha_T$ fit.}
    \label{fig:pvalue_PerSyst_TKI}
\end{figure}

The $p$-values from all types of LLH distributions shown from Figs.~\ref{fig:pvalue_Global}--\ref{fig:pvalue_PerSyst_TKI} are above the threshold of 0.05, which indicate the data is consistent with the prediction and uncertainties.

%% file: Sections/Appendix_LCurve.tex
\section{``L-curve" optimization of regularization strengths}
\label{sec:appd_lcurve}

\begin{figure}[htbp]
    \centering
    \includegraphics[height=0.22\textwidth]{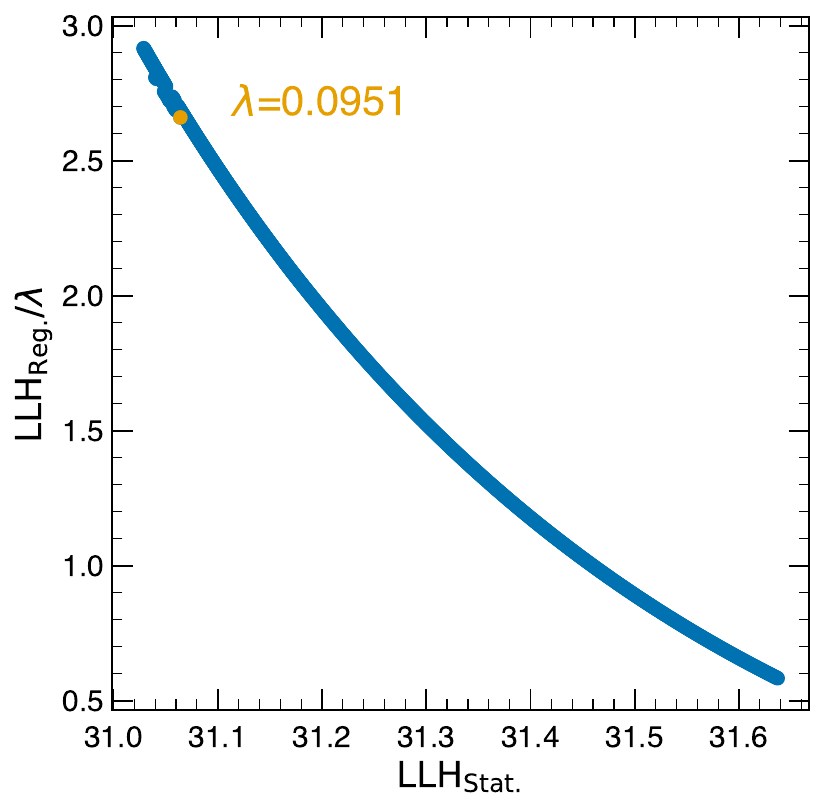}
    \includegraphics[height=0.22\textwidth]{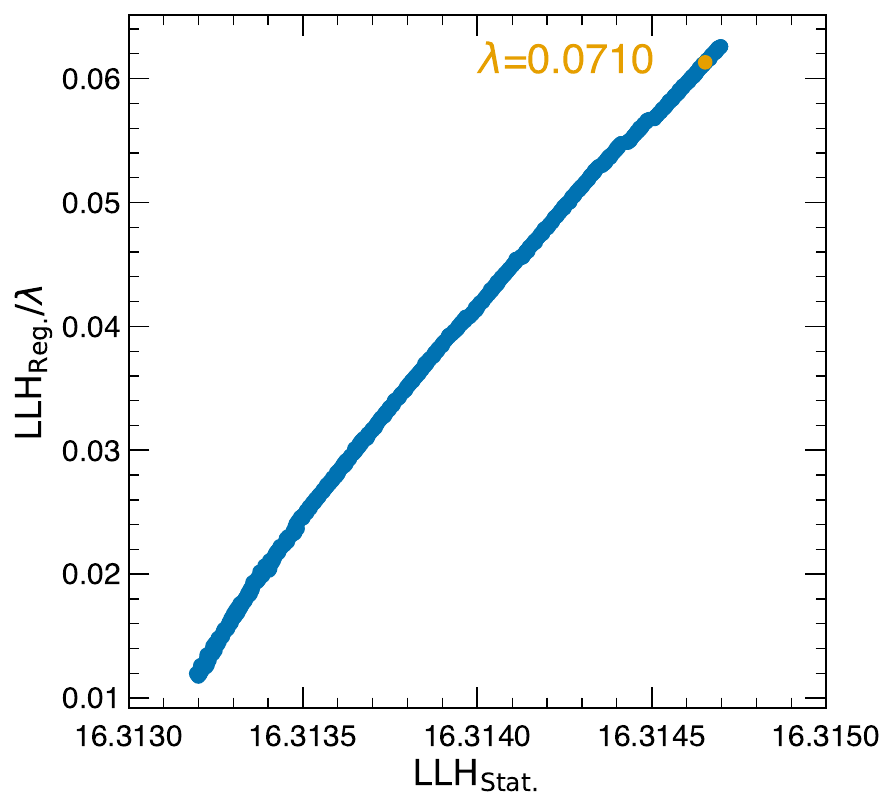}
    \caption{$\mathrm{LLH}_{\mathrm{Reg.}}/\lambda$ versus $\mathrm{LLH}_{\mathrm{Stat.}}$ used in the ``L-curve" optimization for the angular (left) and TKI (right) variables. Orange points are the ones that maximize curvature.}
    \label{fig:TemplateRegCovLCurve}
\end{figure}

In Sec.~\ref{sec:xsecextraction:fitting}, we described the regularization of template parameters to break the degeneracy in the best-fit results.
The strength of the regularization is defined by $\lambda_{\mathrm{Reg.}}$ and can be optimized from the ``L-curve"~\cite{doi:10.1137/1034115}.
The two LLH values, $\mathrm{LLH}_{\mathrm{Reg.}}/\lambda$ and $\mathrm{LLH}_{\mathrm{Stat.}}$ are plotted for each value of $\lambda_{\mathrm{Reg.}}$ used in the fit.
An increase of the former represents how much regularization is introduced in the fit and the latter indicates the amount of bias in the post-fit prediction.
The optimal value of $\lambda_{\mathrm{Reg.}}$ can be found by selecting the value with maximum curvature, where the curvature is evaluated in the distribution of $\mathrm{LLH}_{\mathrm{Reg.}}/\lambda$ and $\mathrm{LLH}_{\mathrm{Stat.}}$.
Figure~\ref{fig:TemplateRegCovLCurve} shows the summary of the optimization with the chosen values used in the analysis; $\lambda_{\mathrm{Reg.}}$ = 0.0951 (0.0710) for angular (TKI) variables.

%% file: Sections/Appendix_IndvFitComparison.tex
\section{Comparison to uncorrelated fit results}
\label{sec:appd_comparison_to_indvfit}

\begin{figure*}
    \centering
    \includegraphics[width=0.90\textwidth]{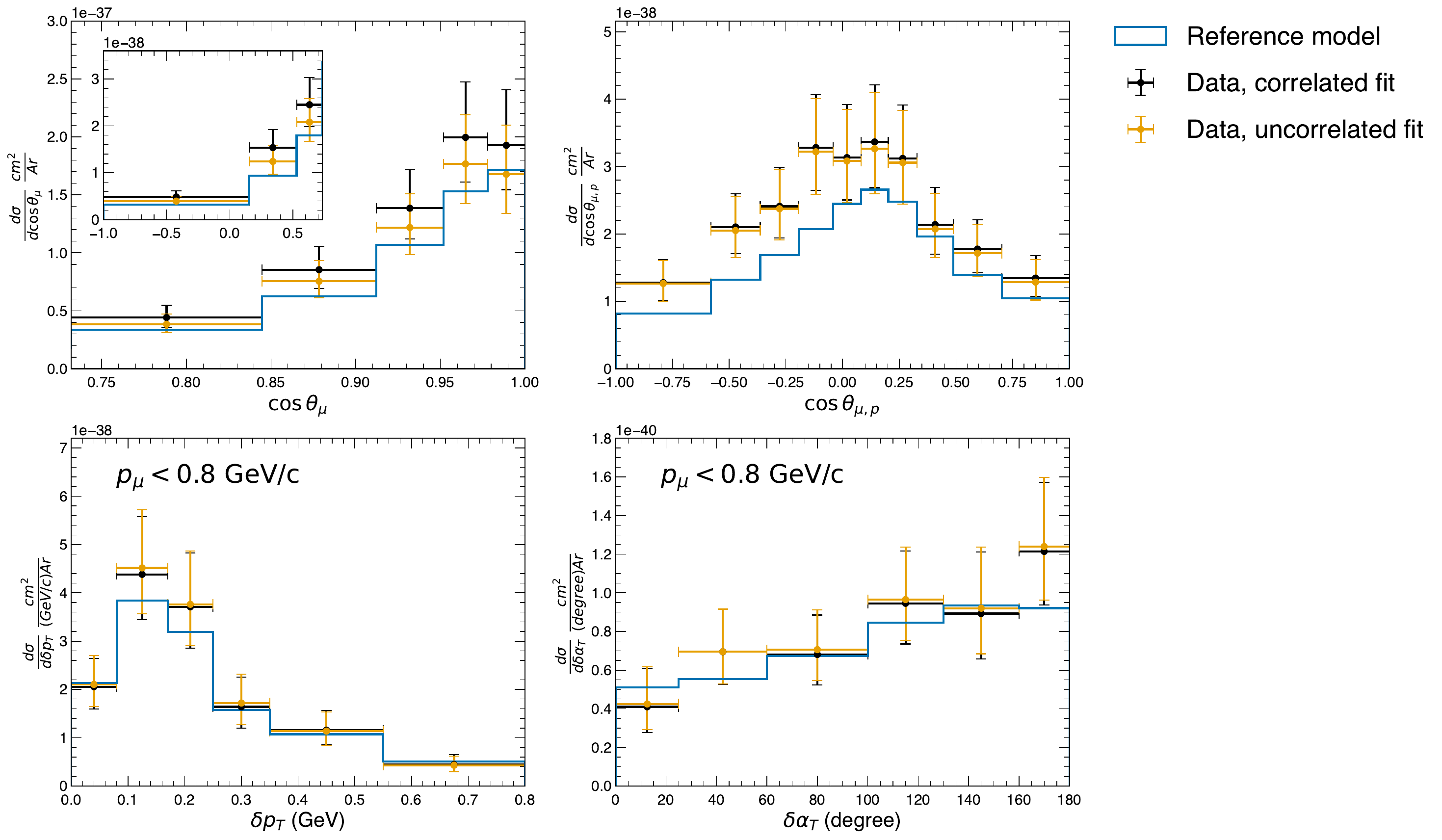}
    \caption{The cross sections extracted from ``uncorrelated" fits (orange marker) are compared to results reported in the top plots of Fig.~\ref{fig:xsec_FineBin_GENIEBreakDown} which used the simultaneous (``correlated") fits described in Sec.~\ref{sec:xsecextraction:fitting} (black marker). The prediction from the \GENIE reference model is also shown (solid histogram).}
    \label{fig:ComparisonToIndvFit}
\end{figure*}

As described in Sec.~\ref{sec:xsecextraction:fitting}, two one-dimensional angular or TKI variables are fit simultaneously (``correlated" fit) using the statistical covariance matrix (Eq.~\ref{eq:LLHStatCov}).
We also performed an ``uncorrelated" fit for each one-dimensional variable, replacing $\mathrm{LLH}_{\mathrm{Stat.}}$ with a ``Barlow--Beeston" log-likelihood ($\mathrm{LLH}_{\mathrm{BB}}$), an improved version of $\mathrm{LLH}_{\mathrm{Poisson}}$ that takes into account the finite simulation statistics~\cite{BARLOW1993219}:
\begin{align}
\label{eq:BBLLHStat}
\mathrm{LLH}_{\mathrm{BB}} &= 2 \sum_{\text{Reco}} \left( \beta M - D + D \log{\frac{D}{\beta M}} + \frac{\beta^{2}-1}{2\sigma^{2}_{M}}  \right)
\end{align}, where
$\beta$ in Eq.~\ref{eq:BBLLHStat} is a scaling factor for each $M$ that minimizes LLH for a given parameter set:
\begin{align}
\label{eq:BBLLHStat_Beta}
\beta = \frac{-(M \sigma^{2}_{M}-1) + \sqrt{(M \sigma^{2}_{M}-1)^{2} + 4M^{2}\sigma^{2}_{M}}}{2}
\end{align}.
Figure~\ref{fig:ComparisonToIndvFit} compares the extracted cross sections from individual fits to the results from the simultaneous fits (i.e., data points in Fig.~\ref{fig:xsec_AnaBin_GENIEBreakDown}).
The extracted cross sections are comparable in terms of both central value and the uncertainties between the two methods.

%% file: Sections/Appendix_XsecFineBins.tex
\section{Cross-section predictions in finer binning}
\label{sec:appd_xsec_FineBin}

\begin{figure*}
    \centering
    \includegraphics[width=0.90\textwidth]{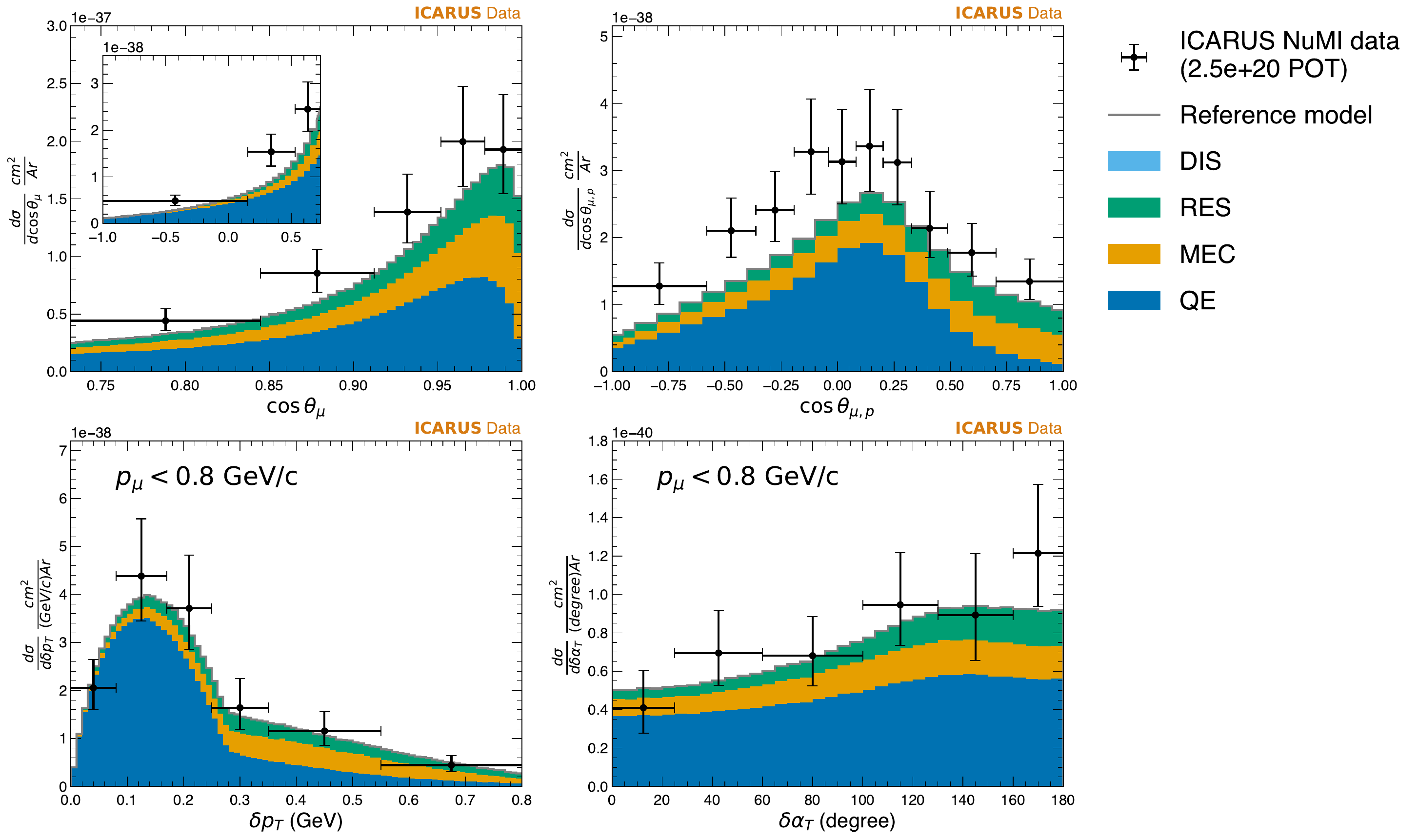}
    \caption{The extracted cross sections (black marker) compared to reference GENIE model (\texttt{AR23\_20i}) prediction broken down into interaction mode (QE, MEC, RES and DIS). As described in Sec.~\ref{sec:RecoSel:SigDef} and \ref{sec:RecoSel:EvSel}, the signal events used to extract cross sections as a function of TKI variables (bottom) are restricted to have muon momentum below $0.8$\,GeV/c.}
    \label{fig:xsec_FineBin_GENIEBreakDown}
\end{figure*}

\begin{figure*}
    \centering
    \includegraphics[width=0.90\textwidth]{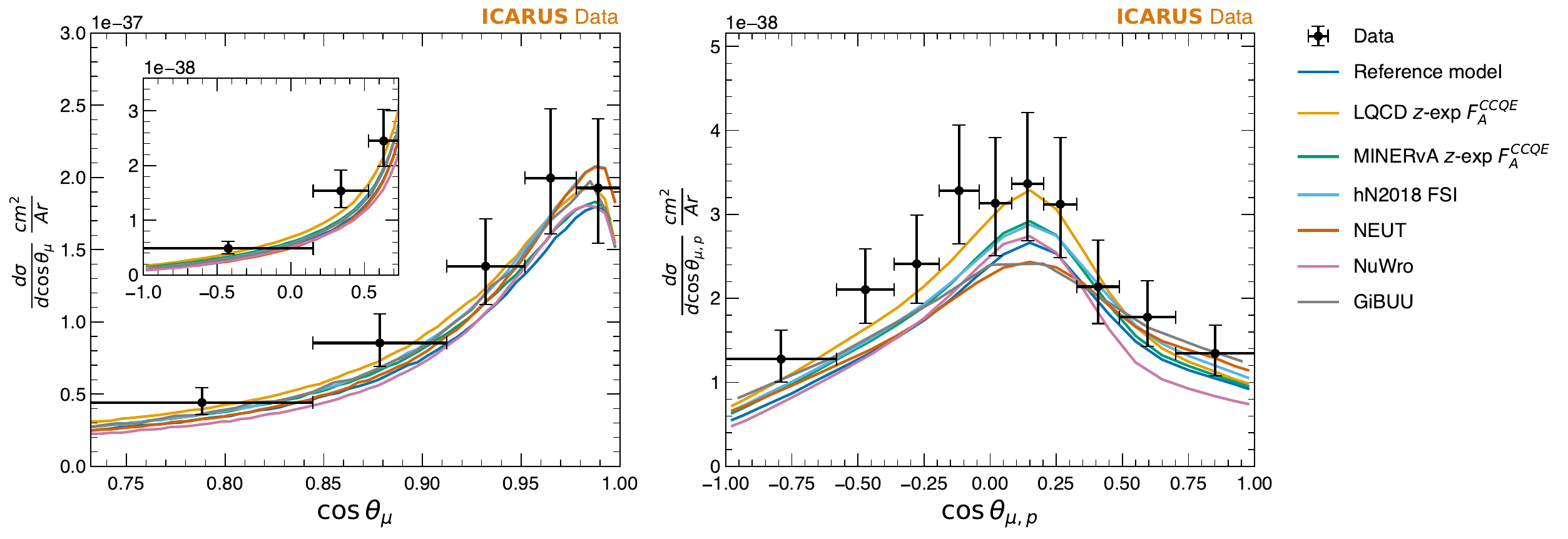}
    \includegraphics[width=0.90\textwidth]{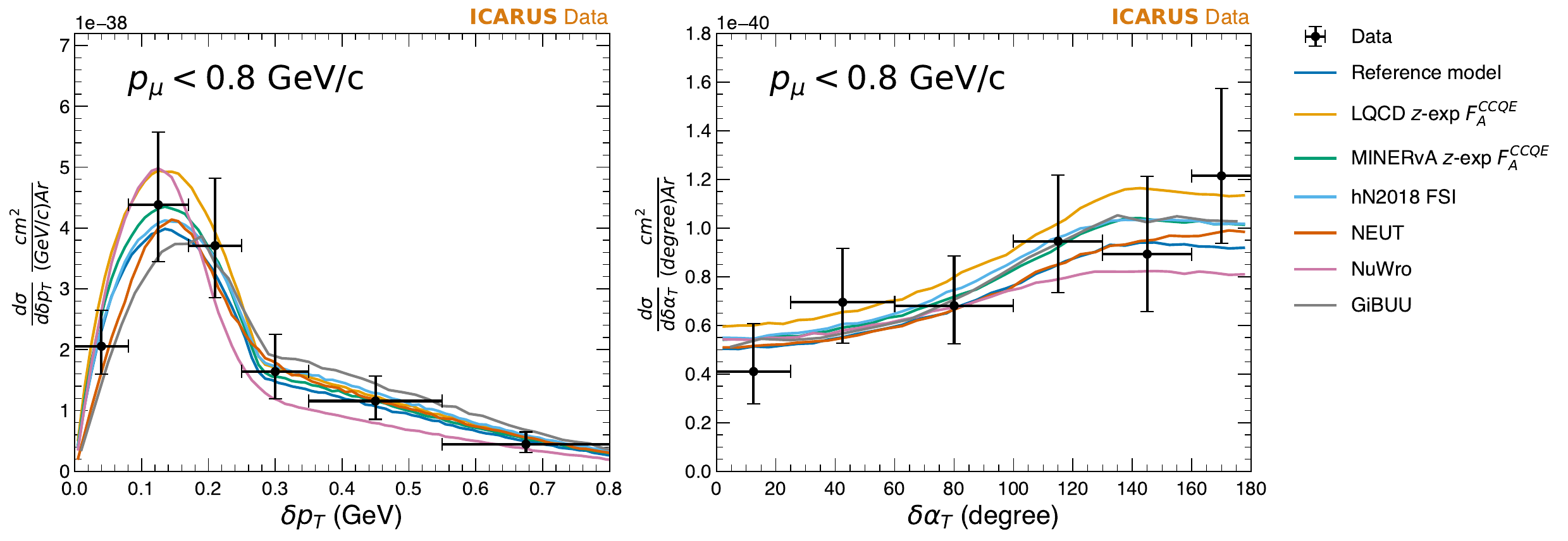}
    \caption{The extracted cross section measured in $\cos{\theta_\mu}$ and $\cos{\theta_{\mu,p}}$ ($\delta p_{T}$ and $\delta \alpha_{T}$) are shown in black marker on the top (bottom) plots.  Various model predictions are shown in finer binning.}
    \label{fig:xsec_FineBin}
\end{figure*}

In this section, we present a similar set of plots to those in Figs.~\ref{fig:xsec_AnaBin_GENIEBreakDown}--\ref{fig:xsec_AnaBin_TKI_With_RatioToGENIE}, but we employ a finer binning for the predictions to better illustrate the underlying shape information.
In Fig.~\ref{fig:xsec_FineBin_GENIEBreakDown} is the data points (black marker) are identical to Fig.~\ref{fig:xsec_AnaBin_GENIEBreakDown}, and the reference model prediction is generated using a finer binning.
Similar to Fig.~\ref{fig:xsec_AnaBin_Angular_With_RatioToGENIE} and \ref{fig:xsec_AnaBin_TKI_With_RatioToGENIE}, Fig.~\ref{fig:xsec_FineBin} shows the extracted cross section (black marker) compared with various model predictions described in Sec.~\ref{sec:ResultAndComparison} but in a finer binning.

